%% file: ms.tex
\documentclass{aa}

\usepackage{graphicx}
\usepackage{natbib}
\bibpunct{(}{)}{;}{a}{}{,} 
\usepackage{txfonts}
\usepackage{xcolor}
\usepackage{listings}
\newcommand{\am}{\texttt{AMPEL}\xspace}

\newcommand{\todo}[1]{}
\newcommand{\new}[1]{#1}
\newcommand{\nnew}[1]{#1}
\newcommand{\rep}[1]{{#1}}

\title{Transient processing and analysis using AMPEL: Alert Management, Photometry and Evaluation of Lightcurves}
\titlerunning{AMPEL. Alert Management, Photometry and Evaluation of Lightcurves}

\author{J.~ Nordin\inst{\ref{inst1}}
  \and V.~Brinnel\inst{\ref{inst1}}
  \and J.~van~Santen\inst{\ref{inst2}}
  \and M.~Bulla\inst{\ref{inst3}}
  \and U.~Feindt\inst{\ref{inst3}}
  \and A.~Franckowiak\inst{\ref{inst2}}
  \and C.~Fremling\inst{\ref{inst4}}
  \and A.~Gal-Yam\inst{\ref{inst5}}
  \and M.~Giomi\inst{\ref{inst1}}
  \and M.~Kowalski\inst{\ref{inst1},\ref{inst2}}
  \and A.~Mahabal\inst{\ref{inst4},\ref{inst6}}
  \and N.~Miranda\inst{\ref{inst1}}
  \and L.~Rauch\inst{\ref{inst2}}
  \and S.~Reusch\inst{\ref{inst1}}
  \and M.~Rigault\inst{\ref{inst7}}
  \and S.~Schulze\inst{\ref{inst5}}
  \and J.~Sollerman\inst{\ref{inst3},\ref{inst8}}
  \and R.~Stein\inst{\ref{inst2}}
  \and O.~Yaron\inst{\ref{inst5}}
  \and S.~van~Velzen\inst{\ref{inst9}}
  \and C.~Ward\inst{\ref{inst9}}
}

\institute{Institute of Physics, Humboldt-Universit\"at zu Berlin, Newtonstr. 15, 12489 Berlin, Germany\label{inst1} \and Deutsches Elektronen-Synchrotron, D-15735 Zeuthen, Germany\label{inst2} \and The Oskar Klein Centre, Department of Physics, Stockholm University, AlbaNova, SE-106 91 Stockholm, Sweden \label{inst3}
\and Division of Physics, Mathematics, and Astronomy, California Institute of Technology, Pasadena, CA 91125, USA \label{inst4}
\and Department of Particle Physics and Astrophysics, Weizmann Institute of Science 234 Herzl St., Rehovot, 76100, Israel \label{inst5}
\and Center for Data Driven Discovery, California Institute of Technology, Pasadena, CA 91125, USA \label{inst6}
\and Université Clermont Auvergne, CNRS/IN2P3, Laboratoire de Physique de Clermont, F-63000 Clermont-Ferrand, France \label{inst7}
\and Department of Astronomy, Stockholm University, AlbaNova, SE-106 91 Stockholm, Sweden \label{inst8}
\and Department of Astronomy, University of Maryland, College Park, MD 20742, USA \label{inst9}
}

\begin{document}

\abstract {Both multi-messenger astronomy and new high-throughput wide-field surveys require flexible tools for the selection and analysis of astrophysical transients.} {We here introduce the Alert Management, Photometry and Evaluation of Lightcurves (\am) system, an analysis framework designed for high-throughput surveys and suited for streamed data. \rep{\am combines the functionality of an alert broker with a generic framework capable of hosting user-contributed code, that encourages provenance and keeps track of the varying information states that a transient displays. The latter concept includes information gathered over time and data policies such as access or calibration levels.}} {We describe a novel ongoing real-time multi-messenger analysis using \am to combine IceCube neutrino data with the alert streams of the Zwicky Transient Facility (ZTF). We also reprocess the first four months of ZTF public alerts, and compare the yields of more than 200 different transient selection functions to quantify efficiencies for selecting Type Ia supernovae that were reported to the Transient Name Server (TNS).}{We highlight three channels suitable for (1) the collection of a complete sample of extragalactic transients, (2) immediate follow-up of nearby transients and (3) follow-up campaigns targeting young, extragalactic transients. We confirm ZTF completeness in that all TNS supernovae positioned on active CCD regions were detected.}{\am can assist in filtering transients in real time, running alert reaction simulations, the reprocessing of full datasets as well as in the final scientific analysis of transient data. This is made possible by a novel way to capture transient information through sequences of evolving states, and interfaces that allow new code to be natively applied to a full stream of alerts. This text also introduces how users can design their own channels for inclusion in the \am live instance that parses the ZTF stream and the real-time submission of high quality extragalactic supernova candidates to the TNS.}

\maketitle

\section{Introduction}

Transient astronomy has traditionally used optical telescopes to detect variable objects, both within and beyond our Galaxy, with a peak sensitivity for events that vary on \new{weekly or monthly} timescales.\todo{references, maybe 11fe and some early famous ones?}
This field has now entered a new phase in which multi-messenger astronomy allows for near real-time detections of transients through correlations between observations of different messengers. The initial report of GW170817 from LIGO/VIRGO, and the subsequent search and detection of an X-ray/optical counterpart, provides a first, inspiring example of this \citep{2017ApJ...848L..12A}.
Shortly after, the observation of a flaring blazar coincident with a high-energy neutrino detected by IceCube illustrated again the scientific potential of time domain multi-messenger astronomy \citep{2018Sci...361.1378I}. 
Optical surveys now observe the full sky daily, to a depth which encompasses both distant, bright objects and nearby, faint ones. We can thus simultaneously find rare objects, obtain an accounting of the variable Universe, and probe fundamental physics at scales beyond the reach of terrestrial accelerators. Exploiting these opportunities is currently constrained as much by software and method development as by available instruments \citep{2018arXiv180704780A}.

The plans for the Large Synoptic Survey Telescope (LSST) provide a sample scale for high-rate transient discovery.
LSST is expected to  scan large regions of the sky to great depth, with sufficient cadence for more than $10^6$ astrophysical transients to be discovered each night. Each such detection will be immediately streamed to the community as an \emph{alert}. The challenge of distributing this information for real-time follow-up observations is to be solved through a set of \emph{brokers}, which will receive the full data flow and allow end-users to select the small subset that merits further study \citep{2017ASPC..512..279J}. \new{Development first started on} the Arizona-NOAO Temporal Analysis and Response to Events System (ANTARES), which provides a system for real-time characterization and annotation of alerts before they are relayed further downstream \citep{2014SPIE.9149E..08S}. \new{Other current brokers include MARS \footnote{https://mars.lco.global/} and LASAIR \citep{Smith_2019}.}
%
Earlier systems for transient information distribution include the \new{Central Bureau for Astronomical Telegrams (CBAT)}, the Gamma-ray Coordinates Network and the Astronomer's Telegram. \rep{The Catalina Real-Time Transient Survey was deisgned to make transient detections public within minutes of observation \citep{2009ApJ...696..870D,2011BASI...39..387M}.}
More recent developments include the Astrophysical Multimessenger Observatory Network \citep[AMON,][]{2013APh....45...56S}, which provides a framework for real-time correlation of transient data streams from different high-energy observatories, and the Transient Name Server (TNS), which maintains the current IAU repository for potential and confirmed extragalactic transients\footnote{https://wis-tns.weizmann.ac.il/}.

While LSST will come online only in 2022, the Zwicky Transient Facility (ZTF) has been operating since March 2018 \citep{graham}.
ZTF employs a wide-field camera mounted on the Palomar P48 telescope, and is capable of scanning more than $3750$ square degrees to a depth of $20.5$ mag each hour \citep{2019PASP..131a8002B}. This makes ZTF a wider, shallower precursor to LSST, with a depth more suited to spectroscopic follow-up observations. 
%
ZTF observations are immediately transferred to the Infrared Processing \& Analysis Center (IPAC) data center for processing and image subtraction  \citep{2019PASP..131a8003M}. Any significant point source-like residual flux in the subtracted image triggers the creation of an alert. Alerts are serialized and distributed through a Kafka\footnote{https://kafka.apache.org} server hosted at the DiRAC centre at University of Washington \citep{2019PASP..131a8001P}. Each alert contains primary properties like position and brightness, but also ancillary detection information  and higher-level derived values such as the RealBogus score which aims to distinguish real detections from image artifacts \citep{2019PASP..131c8002M}. Full details on the reduction pipeline and alert content can be found in \cite{2019PASP..131a8003M}, while an overview of the information distribution can be found in the top row of Fig.~\ref{fig:system}.
ZTF will conduct two public surveys as part of the US NSF Mid-Scale Innovations Program (MSIP). One of these, the Northern Sky Survey, performs a three-day cadence survey in two bands of the visible Northern Sky.

We here present \am (Alert Management, Photometry and Evaluation of Lightcurves) as a tool to accept, 
process and react to streams of transient data. 
\am contains a broker as the first of four pipeline levels, or 'tiers', 
but complements this with a framework enabling analysis methods to be easily and consistently applied to large data volumes. 
The same set of input data can be repeatedly reprocessed with progressively refined analysis software, while the same algorithms can then also be applied to real-time, archived and simulated data samples. Analysis and reaction methods can be contributed through the implementation of simple \texttt{python} classes, 
ensuring that the vast majority of current community tools can be immediately put to use. 
\am functions as a public broker for use with the public ZTF alert stream, meaning that community members can provide analysis units for inclusion in the real-time data processing.
\nnew{\am also brokers alerts for the private ZTF partnership.
  Selected transients, together with derived properties, are pushed into the GROWTH Marshal \citep{growth} for visual examination, discussion and the potential trigger of follow-up observations.}

This paper is structured as follows: \am requirements are first described in Sec.~\ref{sec:req}, after which the design concepts are presented in Sec.~\ref{sec:nut}, some specific implementation choices detailed in Sec.~\ref{sec:imp} and instructions for using \am are provided in Sec.~\ref{sec:use}. In Sec.~\ref{sec:app} we present sample \am uses: systematic reprocessing of archived alerts to investigate transient search completeness and efficiency, photometric typing and live multi-messenger matching between optical and neutrino data-streams.
The discussion (Sec.~\ref{sec:disc}) introduces the automatic \am submission of high-quality extragalactic astronomical transients to the TNS, from which astronomers can immediately find potential supernovae or AGNs without having to do any broker configuration. 
The material presented here focuses on the design and concepts of \am, and acts as a complement to the software design tools contained in the \am sample repository\footnote{https://github.com/AmpelProject/Ampel-contrib-sample}. We encourage the interested reader to look at this in parallel to this text.
\rep{We describe the \am system using terms where the interpretation might not match that used in other fields. This terminology will be introduced gradually in this text, but is summarized in Table~\ref{tab:terms} for reference.}

\begin{table*}
  \caption{\am terminology}
  \centering
  \label{tab:terms}
  \begin{tabular}{ | l | p{10cm} |}
    \hline
    Term & \am interpretation  \\ \hline \hline
    Transient &  Object with a unique ID provided from a data source and accepted into \am by at least one \emph{channel}. \\
    Datapoint & A single measurement with a specific calibration, processing level etc. \\
    Compound & A collection of \emph{datapoints} (from one or more instruments). \\
    State & A view of a transient object available at some point for some observer. Connects a \emph{compound} with one (or more) \emph{transients}. \\
    Tier & \am is internally divided into four tiers, where each performs different kinds of operations and is controled by a separate scheduler. \\
    Channel & Configuration of requested behaviour at all \am \emph{tiers} supplied by a user (for one science goal). Typically consists of a list of requested \emph{units} together with their run parameters. \\
    Archive & All alert data, also those rejected during live processing, are stored in an archive for \emph{reprocessing}. \\
    ScienceRecord & Records the result of a science computation made based on data available in specific \emph{state}. \\
    TransientView & All information available regarding a specific \emph{transient}. This can include multiple \emph{states}, and any \emph{ScienceRecords} associated with these. \\
    Unit & Typically implemented as \texttt{python} modules, a \emph{unit} allows user contributed code to be directly called during data processing. \emph{Units} at different \emph{tiers} receive different input and are expected to produce different kinds of output. \\
    Journal & A time-ordered log included in each \emph{transient}. \\
    Purge & The transfer of a no longer active \emph{transient} from the live database to external storage. This includes all connected \emph{datapoints}, \emph{states}, \emph{compounds} and \emph{ScienceRecords}. \\
    Live instance & A version of \am processing data in real-time. This includes a number of active \emph{channels}. \\
    Reprocessing & Parsing archived alerts as they would have been received in real-time, using a set of \emph{channels} defined as for a \emph{live instance}. \\
    \hline
    \end{tabular}
\end{table*}

\section{Requirements}\label{sec:req}

Guided by an overarching goal of analyzing data streams, we here lay out the design requirements that shaped the \am development:

\paragraph{Provenance and reproducibility:}
Data provenance encapsulates the philosophy that the origin and manipulation of a dataset should be easily traceable. As data volumes grow, and as astronomers increasingly seek to combine ever more diverse datasets, the concept of data provenance will be of central importance. In this era, individual scientists can be expected neither to master all details of a given workflow, nor to inspect all data by hand. As an alternative, these scientists must instead rely on documentation accompanying the data.
While provenance is a minimal requirement for such analysis, a more ambitious goal is \emph{replayability}. Replaying an archival transient survey offline would involve providing a virtual survey in which the entire analysis chain is simulated, from transient detection to the evaluation of triggered follow-up observations. In essence, this amounts to answering the question: \emph{If I had changed my search or analysis parameters, what candidates would have been selected?}
Because any given transient will only be observed once, replayability is as close to the standard scientific goal of reproducibility as astronomers can get.

\paragraph{Analysis flexibility:}
The next decades will see an unprecedented range of complementary surveys looking for transients through gravitational waves, neutrinos and multiwavelength photons. These will feed a sprawling community of diverse science interests. We would like a transient software framework that is sufficiently flexible to give full freedom in analysis design, while still being compatible with existing tools and software.

\paragraph{Versions of data and software:}
It is typical that the value of a measurement evolves over time, from a preliminary real-time result to final published data.
This is driven both by changes in the quantitative interpretation of the observations, as well as a progressive increase in analysis complexity. The first dimension involves changes such as improved calibration, while the second incorporates, for example, more computationally expensive studies only run on subsets of the data.
So far it has been hard to study the full impact of incremental changes in these two dimensions. \new{To change this requires an end-to-end streaming analysis framework where any combinations of data and software can be conveniently explored.}
A related community challenge is to recognize, reference and motivate continued development of well-written software.

\paragraph{Alert rate:}
Current optical transient surveys such as DES, ZTF, ASAS-SN and ATLAS, as well as future ones (LSST), do or will provide tens of thousands to millions of detections each night. With such scale, it is impossible for human inspection of all candidates, even assuming that artifacts could be perfectly removed\footnote{For optical surveys, a majority of these ``detections'' are actually artifacts induced through the subtraction of a reference image. Machine learning techniques, such as RealBogus for ZTF, are increasingly powerful at separating these from real astronomical transients. However, this separation can never be perfect and any transient program has to weigh how aggressively to make use of these classifications.}.
A simplistic solution to this problem is to only select a very small subset from the full stream, for example a handful of the brightest objects, for which additional human inspection is feasible.
A more complete approach would be based on retaining much larger sets of targets throughout the analysis, from which subsets are complemented with varying levels of follow-up information. As the initial subset selection will by necessity be done in an automated streaming context, the accompanying analysis framework must be able to trace and model these real-time decisions.

\section{\am in a nutshell}\label{sec:nut}

\begin{figure*}
\begin{centering}
\includegraphics[page=1,width=0.85 \textwidth,trim={0 15.cm 0 0},clip]{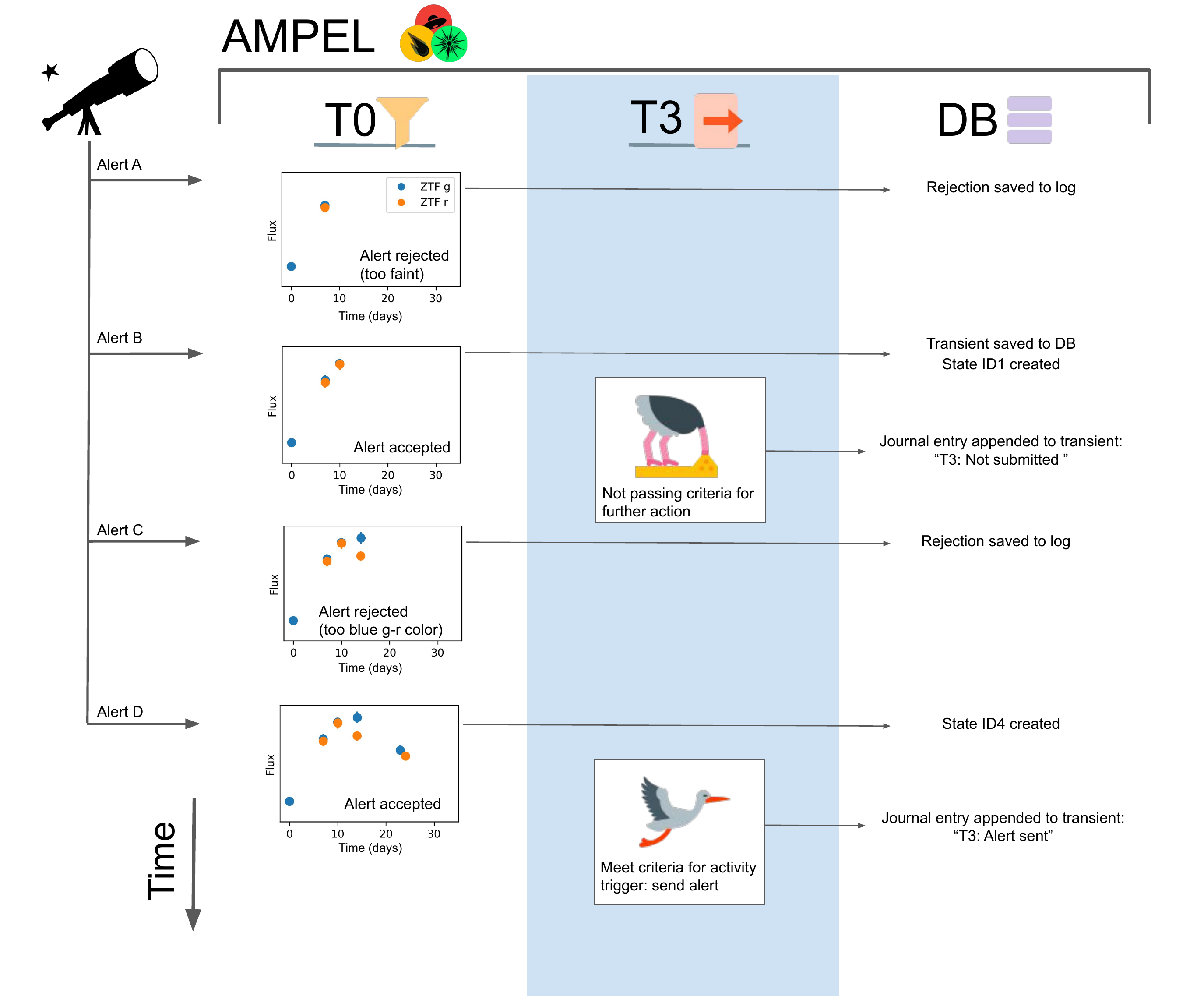}
\caption{\emph{Outline of \am, acting as broker.} Four alerts, A to D, belonging to a unique transient candidate are being read from a stream. In a first step, ``Tier 0'', the alert stream is \emph{filtered} based on alert keywords and catalog matching. Alerts B and D are accepted. In a second step, ``Tier 3'', it is decided which external resources \am should notify. In this example, only Alert D warrants an immediate reaction. The final column shows the corresponding database events. }
\label{fig:brokerlife}
\end{centering}
\end{figure*}

\begin{figure*}
\begin{centering}
\includegraphics[page=1,width=0.85 \textwidth,trim={0 2.cm 0 0},clip]{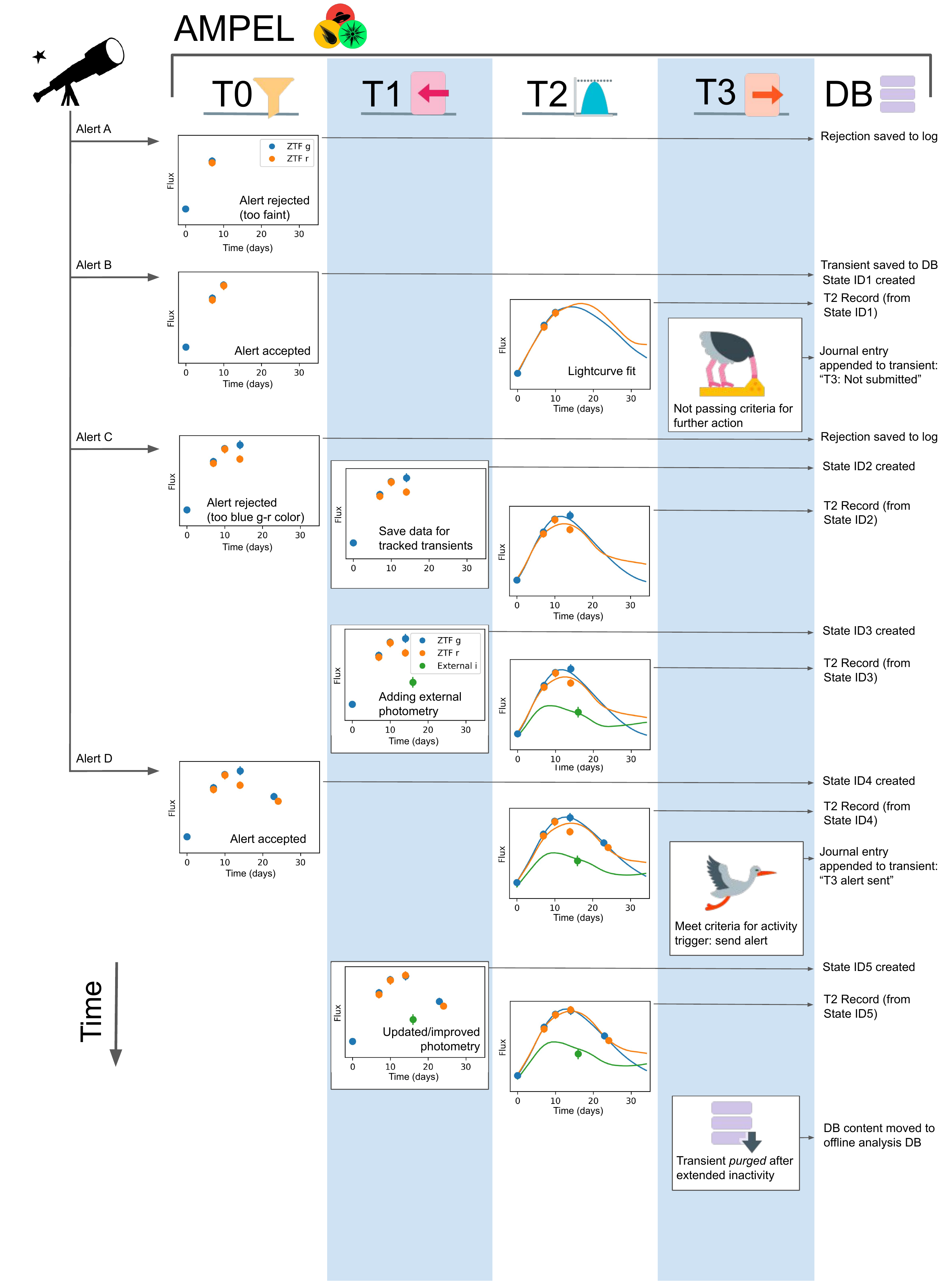}
\caption{\emph{Life of a transient in \am.} Sample behavior at the four tiers of \am as well as the database access are shown as columns, with the left side of the figure indicating when the four alerts belonging to the transient were received.
  \emph{T0:} \rep{The first and third alerts are rejected, while the second and fourth fulfill} the channel acceptance criteria.
  \emph{T1:} \rep{The first T1 panel shows how the data content of an alert which was rejected at the T0 stage but where the transient ID was already known to \am is still ingested into the live DB. The second panel shows an external datapoint (measurement) being added to this transient. The final T1 panel shows how one of the original datapoints is updated. All T1 operations lead to the creation of a new state.}
  \emph{T2:} \rep{The T2 scheduler reacts every time a new state is created and queues the execution of all T2s requested by this channel.} In this case this causes a lightcurve fit to be performed and the fit results stored as a Science Records.
  \emph{T3:} \rep{The T3 scheduler schedules units for execution at pre-configured times. In this example this is a (daily) execution of a unit testing whether any modified transients} warrants a Slack posting (requesting potential further follow-up). The submit criteria are fulfilled the second time the unit is run. In both cases the evaluation is stored in the transient \emph{Journal}, which is later used to prevent a transient to be posted multiple times. Once the transient has not been updated for an extended time a T3 unit \emph{purges} the transient to an external database that can be directly queried by channel owners.
  \emph{Database:} A transient entry is created in the DB as the first alert is accepted. After this, each new datapoint causes a new state to be created. T2 Science Records are each associated with one state. The T3 units return information that is stored in the Journal.
} 
\label{fig:life}
\end{centering}
\end{figure*}

\am is a framework for analyzing and reacting to streamed information, with a focus on astronomical transients. Fulfilling the above design goals requires a flexible framework built using a set of general concepts. These will be introduced in this section, accompanied by examples based on optical data from ZTF. \new{
  The ``life'' of a transient in \am is in parallel  outlined in Figs.~\ref{fig:brokerlife} and~\ref{fig:life}. These further illustrate many of the concepts introduced in this section. Fig.~\ref{fig:brokerlife} shows \am used as a straightforward alert broker, while Fig.~\ref{fig:life} includes many of the additional features that make \am into a full analysis framework.
}

The core object in \am is a \emph{transient}, a single object identified by a creation date and \new{typically} a region of origin in the sky. Each transient is linked to a set of \emph{datapoints} that represent individual measurements\footnote{Note that this is a many-to-many connection; multiple transients can be connected to the same datapoint due to e.g. positional uncertainty. Datapoints can also originate from different sources.}. Datapoints can be added, updated or marked as bad. \rep{Datapoints are never removed.} Each datapoint can be associated with tags indicating e.g. any masking or proprietary restrictions. Transients and datapoints are connected by \emph{states}, where a state references a \emph{compound} of datapoints. A state represents a view of a transient available at some time and for some observer. For an optical photometric survey, a compound can be directly interpreted as a set of flux measurements or a lightcurve.

\vspace{1em}
  \emph{Example:
  A ZTF alert corresponds to a potential transient. Datapoints here are simply the photometric magnitudes reported by ZTF, \rep{which in most cases consists of a recent detection and a history of previous detections or non-detections at this position}. When first inserted, a transient has a single state with a compound consisting of the datapoints in the initial alert. \rep{Should a new alert be received with the same ZTF ID, the new datapoints contained in this alert  are added to the collection and a new state is created containing both previous and new data.} Should the first datapoint be public but the second datapoint be private, only users with proper access will see the updated state. }
\vspace{1em}

Using \am means creating a \emph{channel}, corresponding to a specific science goal, which prescribes behavior at four different stages, or \emph{tiers}.
\new{What tasks should be performed at what tier can be determined} by answers to the questions: ``Tier 0: \emph{What are the minimal requirements for an alert to be interesting?}'', ``Tier 1:  \emph{\new{Can datapoints be changed by events external to the stream?}}'', ``Tier 2:  \emph{What calculations should be done on each of the candidates states?}'', ``Tier 3: \emph{What operations should be done at timed intervals or on populations of transients?}''\footnote{Timed intervals include very high frequencies or effectively real-time response channels.}

\begin{itemize}
  \item Tier 0 (T0) filters the full alert stream to only include potentially interesting candidates. This tier thus works as a data broker: objects that merit further study are selected from the incoming alert stream. However, unlike most brokers, accepted transients are inserted into a database (DB) of active transients rather than immediately being sent downstream. \rep{All alerts, also those rejected, are stored in an external archive DB.} Users can either provide their own algorithm for filtering, or configure one of the filter classes \new{already available} according to their needs. 
\end{itemize}

\vspace{1em}
  \emph{Example T0:
    The simple \am channel ``BrightAndStable'' looks for transients with at least three well behaved detections (few bad pixels and reasonable subtraction FWHM) and not coincident with a Gaia DR2 star-like source. This is implemented through a \texttt{python} class \texttt{SampleFilter} that operates on an alert and returns either a list of requests for follow-up (T2) analysis, if selection criteria are fulfilled, or \texttt{False} if they are not. \am will test every ZTF alert using this class, and all alerts that pass the cut are added to a DB containing all active transients. The transient is there associated with the channel ``BrightAndStable''. }
\vspace{1em}

\begin{itemize}
\item Tier $1$ (T1) is largely autonomous and exists in parallel to the other tiers. \rep{T1 carries out duties of assigning datapoints and T2 run requests to transient states.} Example activities include completing transient states with datapoints that were present in new alerts but where these were not individually accepted by the channel filter (e.g., in the case of lower significance detections at late phases), as well as querying an external archive for updated calibration \new{or adding photometry from additional sources}. \rep{A T1 unit could also parse previous alerts at or close to the transient position for old data to include with the new detection.}
\end{itemize}

\begin{itemize}
  \item Tier $2$ (T2) derives or retrieves additional transient information, and is always connected to a state and stored as a \emph{ScienceRecord}.
    T2 units either work with the empty state, relevant for e.g. catalog matching that only depends on the position, or they depend on the datapoints of a state to calculate new, derived transient properties. In the latter case, the T2 task will be called again as soon as a new state is created.  This could be due both to new observations or, for example, updated calibration of old datapoints. Possible T2 units include lightcurve fitting, photometric redshift estimation, machine learning classification, and catalog matching.
    \end{itemize}

\vspace{1em}
  {\emph{Example T2: For an optical transient, a state corresponds to a lightcurve and each photometric observation is represented by a datapoint. A new observation of the transient would extend the lightcurve and thus create a new state.``BrightAndStable'' requests a third order polynomial fit  for each state  using the \texttt{T2PolyFit} class. The outcome, in this case polynomial coefficients, are saved to the database.
  }}
\vspace{1em}

\begin{itemize}
  \item Tier $3$ (T3), the final \am level, consists of \emph{schedulable} actions. While T2s are initiated by events (the addition of new states), T3 units are executed at pre-determined times. These can range from yearly data dumps, to daily updates or to effectively real-time execution every few seconds. T3 processes access data through the \emph{TransientView}, which concatenates all  information regarding a transient. This includes both states and ScienceRecords that are accessible by the channel.
\rep{T3s iterate through all transients of a channel which were updated since a previous timestamp (either the last time the T3 was run or a specified time-range)}. This allows for an evaluation of multiple ScienceRecords and comparisons between different objects (such as any kind of population analysis). One typical case is the ranking of candidates which would be interesting to observe on a given night.  \nnew{T3 units include options to push and pull information to and from for example the TNS, web-servers and collaboration communication tools such as \texttt{Slack}\footnote{https://slack.com}.}
\end{itemize}

\vspace{1em}
  \emph{Example T3:
The science goal of ``BrightAndStable'' is to observe transients with a steady rise. At the T3 stage the channel therefore loops through the TransientViews, and examines all T2PolyFit science records for fit parameters which indicate a lasting linear rise. Any transients fulfilling the final criteria trigger an immediate notification sent to the user. This test is scheduled to be performed at 13:15 UTC each day.                }                       
\vspace{1em}

\section{Implementation}\label{sec:imp}

\begin{figure*}
\includegraphics[width=1.0 \textwidth,trim={0 1cm 0 3cm},clip]{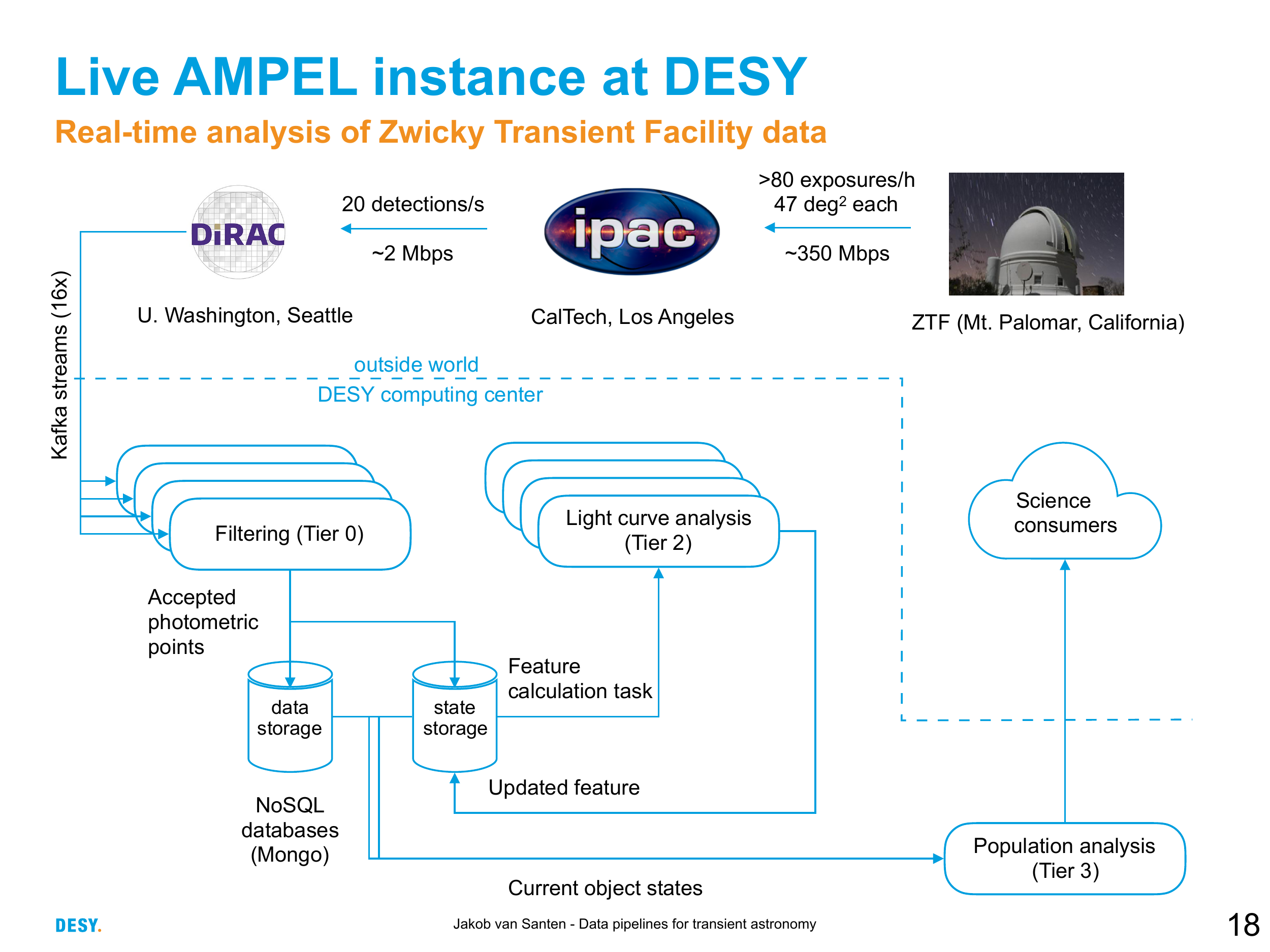}
\caption{\am schematic for the live processing of ZTF alerts. \emph{External events, above dashed lines:} This includes ZTF observations, processing and the eventual alert distribution through the DiRAC centre. Finally, science consumers external to \am receive output information. This can include both full transient display frontends as well as alerts through e.g. TNS or GCN. \emph{Internal processing, below dashed line:} A set of parallel alert processors examine the incoming Kafka Stream (Tier 0). Accepted alert data is saved into a collection, while states are recorded in another. A light curve analysis (Tier 2) is performed on all states. The available data, including the Tier 2 output, is examined in a Tier 3 unit that selects which transients should be passed out. This particular use case does not contain a Tier 1 stage.}
\label{fig:system}
\end{figure*}

We here expand on a selection of implementational aspects. An overview of the live instance processing of the ZTF alert stream can be found in Fig.~\ref{fig:system}.

\paragraph{Modularity and Units}
Modularity is achieved through a system of \emph{units}. These are python modules that can be incorporated with \am and directly be applied to a stream of data. Units are inherited from abstract classes that regulate the input and output data format, but have great freedom in implementing what is done with the input data. The tiers of \am are designed such that each requires a specific kind of information: At Tier $0$ the input is the raw alert content, at Tier $2$ a transient state, and at Tier $3$ a transient view. The system of base classes allows \am to provide each unit with the required data. In a similar system, each unit is expected to provide output data (results) in a specific format to make sure this is stored appropriately: At Tier $0$ the expected output is a list of Tier $2$ units to run at each state for accepted transients (\texttt{None} for rejected transients). At Tier $2$ output is a science record (dictionary) which in the DB is automatically linked to the state from which it was derived. The T3 output is not state-bound, but is rather added to the transient \emph{journal}, a time-ordered history accompanying each transient.
Modules at all tiers can make direct use of well developed libraries such as \texttt{numpy} \citep{numpy}, \texttt{scipy} \citep{scipy}, and \texttt{astropy} \citep{astropy:2013,astropy:2018}. 
Developers can choose to make their contributed software available to other users, and gain recognition for functional code, or keep them private. 
The modularity means that users can independently vary the source of alerts, calibration version, selection criteria and analysis software.\todo{Mention unit tests?} 

\paragraph{Schemas and \am shapers}
Contributed units will be limited as long as they have to be tuned for a specific kind of input, e.g., ZTF photometry. Eventually, we hope that more general code can be written through the development of richer schemas for astronomical information based on which units can be developed and immediately applied to different source streams.
The International Virtual Observatory Alliance (IVOA) initiated the development of the \texttt{VOEvent} standard with this purpose\footnote{http://www.ivoa.net/documents/VOEvent/20110711/REC-VOEvent-2.0.pdf}. Core information of each event is to be mapped to a set of specific tags (such as \emph{Who}, \emph{What}, \emph{Where}, \emph{When}), stored in an \texttt{XML} document.
\texttt{VOEvents} form a starting point for this development \citep[see e.g.][]{2009ASPC..411..115W}, but more work is needed before a general T2 unit can be expected to immediately work on data from all sources. As an intermediate solution, \am employs \emph{shapers} that can translate source-specific parameters to a generalized data format that all units can rely on. 
While the internal \am structure is designed for performance and flexibility, it is easy to construct T3 units that export transient information according to, for example, \texttt{VOEvent} or \texttt{GCN} specifications.

\paragraph{The archive}
Full replayability requires that all alerts are available at later times. While most surveys are expected to provide this, we keep local copies of all alerts until other forms of access are guaranteed.

\paragraph{Catalogs, Watch-lists and ToO triggers}
Understanding astronomical transients frequently requires matches to known source catalogs. \am currently provides two resources to this end. A set of large, pre-packaged catalogs can be accessed using \texttt{catsHTM}, including the Gaia DR2 release \citep{1538-3873-130-989-075002}. As a complement, users can upload their own catalogs using \texttt{extcats}\footnote{https://github.com/MatteoGiomi/extcats} for either transient filtering or to annotate transients with additional information. \texttt{extcats} is also used to create \emph{watch-lists} and \emph{ToO channels}.
Watchlists are implemented as a T0 filter that matches the transient stream with a contributed extcat catalog. 
A ToO channel has a similar functionality, but employs a dynamic \texttt{extcat} target list where a ToO trigger immediately adds one or more entries to the matchlist. The stream can in this case initially be replayed from some previous time (a \emph{delayed T0}), which allows preexisting transients to be consistently detected.

\paragraph{The live database} The live transient DB is built using the NoSQL \texttt{MongoDB}\footnote{https://docs.mongodb.com/manual/} engine. The flexibility of not having an enforced schema allows \am to integrate varying alert content and give full freedom to algorithms to provide output of any shape.
The live \am instance is a closed system that users cannot directly interact with, and contributed units do not directly interact with the DB. Instead, the \am core system manages interactions through the alert, state and transient view objects introduced above\footnote{Eventually, daily snapshot copies of the DB will be made available for users to interactively examine the latest transient information without being limited with what was reconfigured to be exported.}.
Each channel also specifies conditions for when a transient is no longer considered ``live''. At this point it is \emph{purged}: extracted from the live DB together with all states, computations and logs, and then inserted into a channel specific offline DB which is provided to the channel owner.

\new{
\paragraph{Horizontal scaling}
\am is designed to be fully parallelizable. The DB, the alert processors and tier controllers all scale horizontally such that additional workers can be added at any stage to compensate for changes to the workload. Alerts can be processed in any order, i.e. not necessarily in time-order.
}

\paragraph{\am instances and containers}
An \am instance is created through combining tagged versions of core and contributed units into a Docker \citep{Docker} image, which is then converted to the Singularity \citep{Singularity} format for execution by an unprivileged user.
The final product is a unique ``container'' that is immutable and encapsulates
the \am software, contributed units and their dependencies. These can be reused and referenced for later work, even if the host environment changes significantly. The containers
are coordinated with a simple orchestration tool\footnote{https://github.com/AmpelProject/singularity-stack} that exposes an interface similar to Docker's ``swarm mode.''
Previously-deployed \am versions are stored, and can be run off-line on any sequence of archived or simulated alerts.
Several instances of \am  might be active simultaneously, with each processing either a fraction of a full live-stream, or some set of archived or simulated alerts. Each works with a distinct database. The current ZTF alert flow can easily be parsed by a single instance, called \emph{the live instance}. 
A full \am analysis combines this active parsing and reacting to the live streams with subsequent or parallel runs in which the effects of the channel parameters can be systematically explored.

\paragraph{Logs and provenance}
\am contains extensive, built-in \emph{logging} functions. All \am units are provided a logger, and we recommend this to be consistently used. Log entries are automatically tagged with the appropriate channel and transient ID, and are then inserted into the DB. These tools, together with the DB content, alert archive and \am container, make provenance straightforward.
The IVOA has initiated the development of a Provenance Data Model (DM) for astronomy, following the definitions proposed by the W3C
\citep{2017ASPC..512..581S}\footnote{http://www.ivoa.net/documents/ProvenanceDM/20181015/PR-ProvenanceDM-1.0-20181015.pdf}. Scientific information is here described as flowing between \emph{agents}, \emph{entities} and \emph{activities}. These are related through causal relations. 
The \am internal components can be directly mapped to the categories of the IVOA Provenance DM: Transients, datapoints, states and ScienceRecords are entities, Tier units are activities and users, \am maintainers, software developers and alert stream producers are agents. A streaming analysis carried out in \am will thus automatically fulfill the IVOA provenance requirements.

\paragraph{Hardware requirements}
The current live instance installed at the DESY computer center in Berlin-Zeuthen consists of two machines, ``Burst'' and ``Transit''. Real time alert processing is done at Burst (32 cores, 96 GB memory, 1 TB SSD) while alert reception and archiving is done at Transit (20 cores, 48 GB memory, 1 TB SSD + medium-time storage).
This system has been designed for extragalactic programs based on the ZTF survey, with a few $ \times 10^5$ alerts processed each night, of which between $0.1$ and $1\%$ are accepted.
Reprocessing large alert volumes from the archive on Transit is done at a mean rate of $100$ alerts per second. As the ZTF live alert production rate is lower than this, and Burst is a more powerful machine, this setup is never running at full capacity.
It would be straightforward to distribute processing of T2 and T3 tasks among multiple machines, but as the expected practical limitation is access to a common database, this is of limited use until extremely demanding units are requested.

\section{Using \am} \label{sec:use}

\subsection{Creating a channel for the ZTF alert stream}\label{sec:create}

The process for creating \am units and channels is fully described in the \texttt{Ampel-contrib-sample} repository\footnote{ https://github.com/AmpelProject/Ampel-contrib-sample }, which also contains a set of sample channel configurations.
The steps to implementing a channel can be summarized as follows:
\begin{enumerate}
  \item Fork the sample repository and rename it \emph{Ampel-contrib-groupID} where \emph{groupID} is a string identifying the contributing science team.  
  \item \new{Create units through populating the t0/t2/t3 sub-directories with python modules. Each is designed through inheriting from the appropriate base class.} 
  \item Construct the repository channels by defining their parameters in two configuration files: \texttt{channels.json} which \new{defines the channel name} and regulates the T0, T1 and T2 tiers, and \texttt{t3\_jobs.json} which determines the schedule for T3 tasks. These can be constructed to make use of \am units present either in this repository, or from other public \am repositories. 
  \item Notify \am administrators. The last step will trigger channel testing and potential edits. After the channel is verified, it will be added in the list of \am contribution units and included in the next image build. The same channel can also (or exclusively) be applied to archived ZTF alerts.\todo{TODO: Appendix with sample channel configuration, incl decent filter parameters. Find out whether entry points should be mentioned. Add a fifth step in running the unit tests. Find ampel desy email address.}
\end{enumerate}

\subsection{Using \am for other streams}

Nothing in the core \am design is directly tied to the ZTF stream, or even optical data. The only source-specific software class is the Kafka client reading the alert stream, and the alert shapers which make sure key variables such as coordinates are stored in a uniform matter. Using a schema-free DB means that any stream content can be stored by \am for further processing. 
A more complex question concerns how to design units usable with different stream sources. As an example, different optical surveys might use different conventions for how to encode which filter was used, and the reference system and uncertainty of reported magnitudes and powerful alert metrics, such as the RealBogus value of ZTF, are unique. Until common standards are developed, classes will have to be tuned directly to every new alert stream.

\section{Initial \am applications}\label{sec:app}

\subsection{Exploring the ZTF alert parameter space}\label{sec:ztfeval}

It has been notoriously challenging to quantify transient detection efficiencies, search old surveys for new kinds of transients, and predict the likely yield from a planned follow-up campaign.\todo{ref, also further sentence about whether a kilonova with a lightcurve like 170817 been observed?} 
We here demonstrate how \am can assist with such tasks.
For this case study we \emph{reprocess} $4$ months of public ZTF alerts using a set of \am filters spanning the parameter space of the main properties of ZTF alerts. The accepted samples of each channel are, in a second step, compared with confirmed Type Ia supernovae (SNe Ia) reported to the TNS during the same period. We can thus examine how different channel permutations differ in detection efficiency, and at what phase each SN Ia was ``discovered''. The base \emph{comparison sample} consists of $134$ normal SNe Ia. The creation of this sample is described in detail in Appendix\ \ref{app:matchlist}.

We processed the ZTF alert archive from June 2nd 2018 (start of the MSIP Northern Sky Survey) to October 1st using $90$ potential filter configurations based on the \texttt{DecentFilter} class.\todo{TODO: Add decent filter appendix?} In total $28667252$ alerts were included.  Each channel exists on a grid constructed by varying the properties described in Table~\ref{tab:options}. We also include $24$ \emph{OR} combinations where the accept criteria of two filters are combined. 
We further consider two additional versions of each filter or filter-combination:
\begin{enumerate}
  \item Transients in galaxies with known active SDSS or MILLIQUAS active nuclei \citep{2015PASA...32...10F,2017A&A...597A..79P} are rejected, and
  \item Transients are required to be associated with a galaxy for which there is a known NED or SDSS spectroscopic redshift $z<0.1$.
\end{enumerate}
In total, this amounts to 342 combinations. 
All of these variants include some version of alert rejection based on coincidence with a star-like object in either PanSTARRS \citep[using the algorithm of][]{2018PASP..130l8001T} or Gaia DR2 \citep{2018A&A...616A...1G}. We also tested channels not including any such rejection, which lead to transient counts around $10^5$ (an order of magnitude greater than with the star-rejection veto). Reprocessing the alert stream in this way took $5$ days even in a non-optimized configuration, demonstrating that \am can process data at the expected LSST alert rate.

\begin{table*}
  \caption{Dominant channel selection variables and potential settings.}
  \centering
  \label{tab:options}
  \begin{tabular}{ | l | p{10cm} |}
    \hline
    Channel property & Options  \\ \hline \hline
    RealBogus & \emph{Nominal:} Require ML score above $0.3$ or \emph{Strong:} above $0.5$ \\
    Detections & More than $[2,4,6,8]$ (any filter) \\
    Alert History & \emph{New:} Not older than $5$ days, \emph{Multi-night:} $4$ to $15$ days, \emph{Persistent:} Older than $8$ days. \\
    Image Quality & \emph{All:} No requirements, \emph{Good:} Limited cuts on e.g. FWHM and bad pixels, \emph{Excellent:} Strong cuts on e.g. FWHM and bad pixels. \\
    Gaia DR2 & \emph{Nominal:} Reject likely stars from Gaia DR2, \emph{Moderate:} only search in small aperture  or \emph{Disabled}. \\
    Star-Galaxy separation & Using PS1 star-galaxy separation \citep{2018PASP..130l8001T} to reject potential stars (\emph{Hard}), likely stars (\emph{Nominal}) or no rejection (\emph{Disabled}). \\
    Match confusion & \emph{Nominal:} Allow candidates close to nearby (confused) sources,  or \emph{Disabled}: reject anything close to stars even if other sources exist. \\
    \hline
    \end{tabular}
\end{table*}

This study is neither complete nor unbiased: A large fraction of the SNe were classified by ZTF, and we know that the real number of SNe Ia observed is much larger than the classified subset. 
Nonetheless it serves both as a benchmark test for channel creation, as well as a starting point for more thorough analysis.
An estimate of the total number of supernovae we expect to be hidden in the ZTF detections can be obtained through the \texttt{simsurvey} code \citep{2019arXiv190203923F}, in which known transient rates are combined with a realistic survey cadence and a set of detection thresholds\footnote{https://github.com/ufeindt/simsurvey}. The predicted number of Type Ia supernovae fulfilling the criteria of one or more of these channels over the same timespan as the comparison sample and with weather conditions matching the observed was found to be $1033$ (average over 10 simulations). \new{\texttt{Simsurvey} also conveniently returns estimates for other supernova types and we find that an additional $276$ Type Ibc, $92$ Type IIn and $377$ Type IIP supernovae are likely to have been observed by ZTF under the same conditions.} The total number of supernovae present in the alert sample is thus estimated to be $1778$.

\begin{figure*}
  \includegraphics[width=0.49 \textwidth]{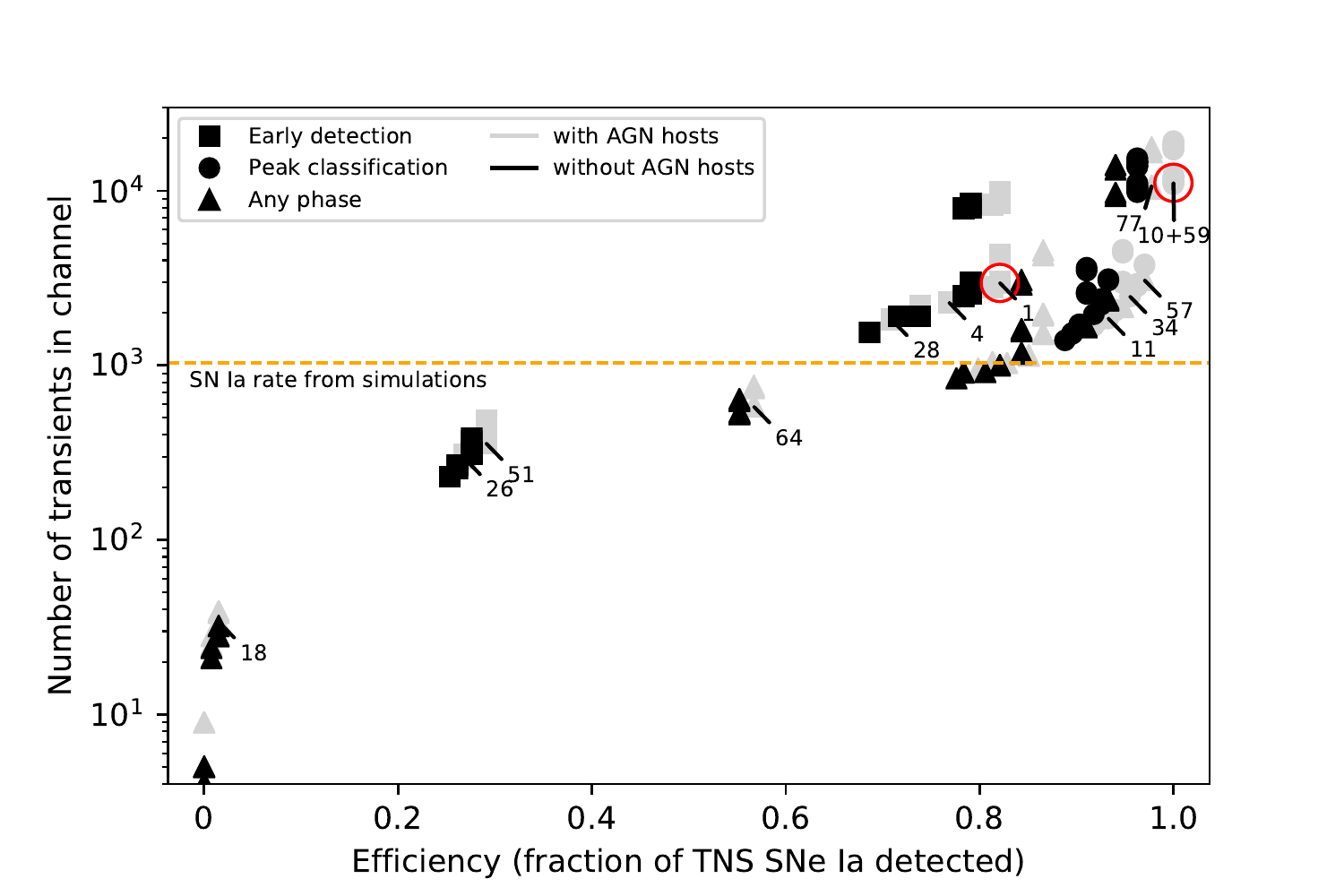}
  \includegraphics[width=0.49 \textwidth]{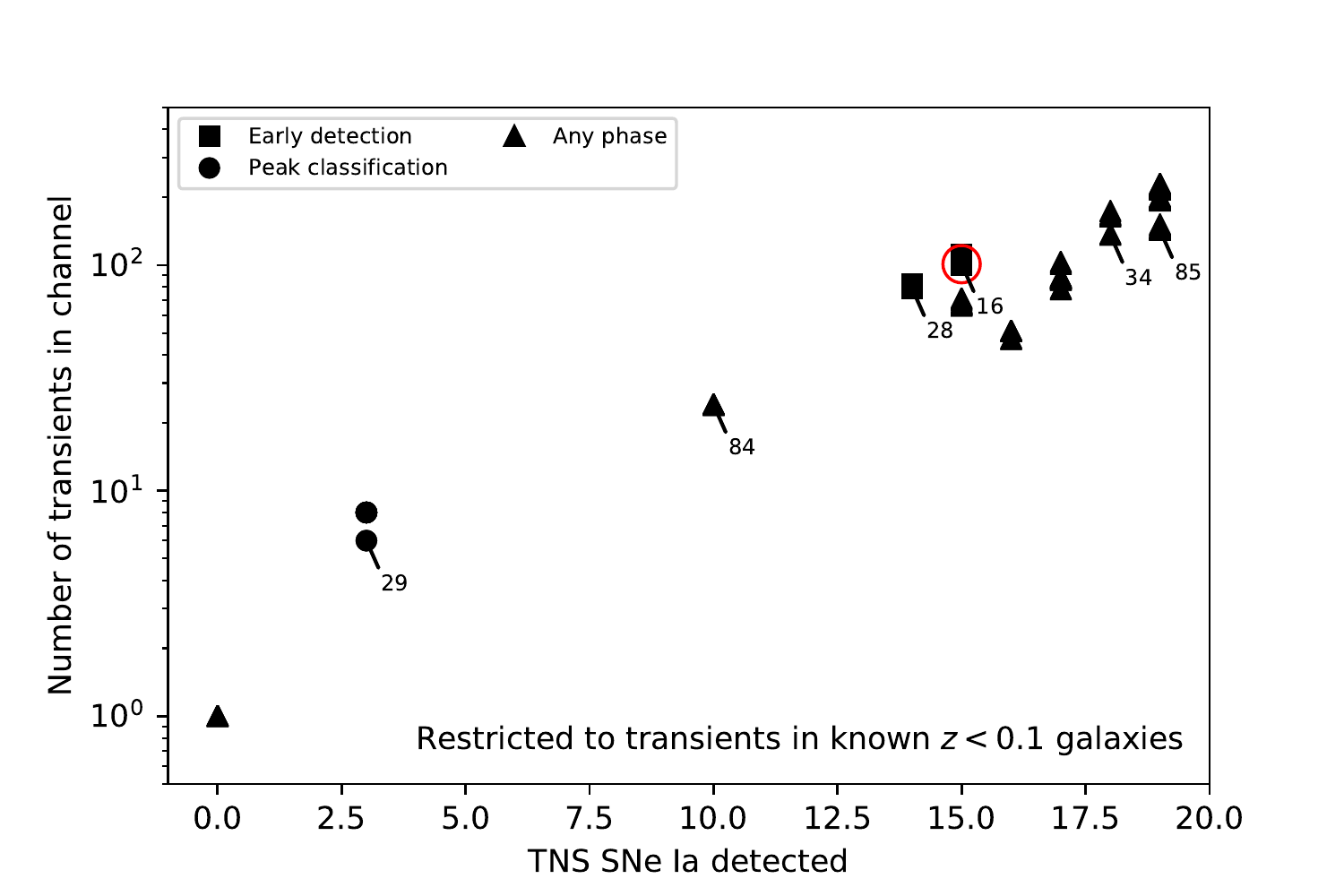}
  \caption{A comparison of the total number of accepted candidates (y-axis) with fraction of the comparison sample SNe Ia detected as x-axis. Symbol shapes indicate the typical phase at which objects in the comparison sample were detected: Channels where more than $25\%$ were detected prior to phase $-10$ are marked as early (squares). If instead more than $95\%$ were detected prior to peak light the channel is defined as suitable for peak classification (circles). Channels not fulfilling either criteria are marked with triangles. \emph{Left panel:} Full channel content. Channels are here divided according to those where transients in galaxies known to host AGN are cut (black) and channels where these are accepted (grey). Compare Table~\ref{tab:detGeneral}. \emph{Right panel:} Comparison of the total number of accepted candidates (y-axis) with the number of comparison sample SNe Ia found, with only candidates linked to a galaxy with known spectroscopic redshift $z<0.1$. All channels reject transients in host galaxies with known AGNs. Compare Table~\ref{tab:detNearby}.  \new{Three channels further discussed in the main text are highlighted (red circles).}
}
  \label{fig:tnscount}
\end{figure*}

The results for channel efficiencies, compared to the total number of accreted transients, can be found in Fig.~\ref{fig:tnscount}. Though we observe the obvious trend that channels with larger coverage of the comparison sample also accept a larger total number of transients, there is also a variation in the total transient counts between configurations that find the same fraction of the comparison sample. Fig.~\ref{fig:tnscount} highlights a subset of the channels as particularly interesting. Selection statistics for these channels can be found in Table~\ref{tab:detGeneral}.
%
For comparison objects with a well defined time of peak light, we also monitor the phase at which it was accepted into each channel. \new{As an estimate for this we use the time of B-band peak light as determined by a SALT lightcurve fit, which is carried out for each candidate at the T2 tier \citep{2014A&A...568A..22B}.}
This information can be used to study how well channels perform in finding early SNe Ia, which constitute a prime target for many supernova science studies. In figure~\ref{fig:tnscount} we therefore mark all channels where more than $25\%$ of all SNe Ia were accepted prior to $-10$ days relative to peak light (``Early detection''). Alternatively, SN Ia cosmology programs often look for a combination of completeness and discovery around lightcurve peak to facilitate spectroscopic classification. Channels not fulfilling the Early detection criteria but where more than $95\%$ of all SNe Ia were accepted prior to peak light are therefore marked as ``Peak classification''. These two simple examples highlight how reprocessing alert streams (reruns) can be used to optimize transient programs, and to estimate yields useful for e.g follow-up proposals.
We also find that a $4\%$ fraction of the comparison sample ($5$ out of $134$) were found in galaxies with documented AGNs, suggesting that programs which prioritize supernova completeness cannot reject nuclear transients with active hosts.

\begin{table*}[htb]
  \caption{\am sample channel parameter settings and rerun statistics. Columns $2$ to $6$ show settings used for parameters in Table~\ref{tab:options}, columns $7$ to $10$ statistics including all targets and columns $11$ to $14$ repeating these when excluding AGN associated candidates. The phase estimates describe the fraction of the matched SNe Ia with a good peak phase estimate that were accepted by the channel either prior to lightcurve peak or prior to $-10$ days with respect to peak light.} 
  \centering
  \label{tab:detGeneral}
  \resizebox{\textwidth}{!}{
    \begin{tabular}{ | l | c | c | c |c |c || c |c |c |c || c |c |c |c |c |}
    \hline
    Channel & RealBogus & Detections & History & Image & Gaia & SNe Ia & Detections  & Phase($<0$) & Phase($<-10$) & SNe Ia & Detections & Phase($<0$) & Phase($<-10$)  \\
    &&&&&& (all) &(all)&(all)&(all)& (no AGN) &(no AGN)&(no AGN)&(no AGN) \\ 

    \hline
    \input{channeltableGeneral.tex}
    \hline
    \end{tabular}
   }
\end{table*}

\new{With \am we are getting closer to one main goal of future transient astronomy -- the immediate, robotic follow-up of the most interesting detections. Facilities such as LCO, the Liverpool Telescope and the Palomar P60 now have the instrumental capabilities for robotic triggers and execution of observations. As the next step towards this we also explore how to select candidates for such automatic programs.}
Figure~\ref{fig:tnscount} (right panel) and Table~\ref{tab:detNearby} show channels where only transients in confirmed nearby galaxies are accepted. While total transient and matched SN Ia counts are much reduced here, all remaining transient candidates can be said with high probability to be both extragalactic and nearby, and thus good candidates for follow-up. Channels such as ``16'' and ``28'' can here be expected to automatically detect multiple early SNe Ia each year and still have small total counts ($160$ and $117$ transients accepted, respectively).

\begin{table*}[htb]
  \caption{\am sample channel parameter settings and rerun statistics for cases when only transients close to $z<0.1$ host galaxies are included. Columns $2$ to $6$ show settings used for parameters in Table~\ref{tab:options}, columns $7$ to $10$ statistics including all targets and columns $11$ to $14$ repeating these when excluding AGN associated candidates. The phase estimates describe the fraction of the matched SNe Ia with a good peak phase estimate that were accepted by the channel either prior to lightcurve peak or to $-10$ days w.r.t. peak.} 
  \centering
  \label{tab:detNearby}
  \resizebox{\textwidth}{!}{
    \begin{tabular}{ | l | c | c | c |c |c || c |c |c |c || c |c |c |c |c |}
    \hline
    Channel & RealBogus & Detections & History & Image & Gaia & SNe Ia & Detections & Phase($<0$) & Phase($<-10$) & SNe Ia & Detections & Phase($<0$) & Phase($<-10$)  \\
    &&&&&& (all) &(all)&(all)&(all)& (no AGN) &(no AGN)&(no AGN)&(no AGN) \\ 

    \hline
    \input{channeltableNearby.tex}
    \hline
    \end{tabular}
   }
\end{table*}

\new{Based on this exploration we highlight three channels:
  \begin{itemize}
    \item \rep{Channel $10+59$, the union of Channels $10$ and $59$ and including AGN galaxies,} is the channel which accepts the least amount of transients while recovering the full comparison sample prior to peak light. We will refer to this as the \emph{``complete''} channel.  
    \item Channel $1$ (including AGN galaxies) strikes a balance between a relatively high completeness ($>80\%$) while detecting transients early and with a limited number of total accepted transients. As will be discussed in Sec.\ref{sec:tns} this channels performs the initial election for the current automatic candidate submission to TNS and is thus referred to as the ``TNS'' channel.   
\item Channel $16$, coupled with only accepting transients in nearby non-AGN host galaxies, provides a very pure selection suitable for automatic follow-up. Consequently, this will be referenced as the \emph{``robotic''} channel. We add ``N'' to the channel number ($16N$) to remind that only transients in nearby ($z<0.1$) galaxies are admitted.
  \end{itemize}
  }

The ``complete'' and ``TNS'' channels differ mainly in that the first accepts transients closer to Gaia sources.

\subsection{Channel content and photometric transient classification}\label{sec:photo}

The previous section examined channels mainly based on the fraction of a known comparison SN Ia sample which was rediscovered. However, as mentioned, the real number of unclassified supernovae (of all types) will be much larger. Every channel will also contain subsets of all other known astronomical variables (e.g. AGNs, variable stars, solar system objects), still unknown astronomical objects and noise. This gap between photometric detections and the number of spectroscopically classified objects will only increase as the number and depth of survey telescopes increase. Developing photometric classification methods is thus  one of the key requisites for the LSST transient program.

ZTF is different in that most transients are nearby and could be classified and the ZTF stream thus provides a way to develop classification methods where the predictions can be verified. As a more immediate application we would like to gain a more general understanding of what transients the \am channels produce.
As a first step in this process we can use the SN Ia template fits introduced in Sec.~\ref{sec:ztfeval} as a primitive photometric classifier. The fits were carried out using a T2 wrapper to the \texttt{SNCOSMO} package\footnote{https://sncosmo.readthedocs.io}. \new{In this case the run configuration only requested the SALT2 SN Ia model to be included, but any transient template could have been requested.} During the stream processing a fit will be done to each state, but we here only analyze the final state fit as we are investigating sample content rather than the evolution of classification accuracy with time (the latter question is more interesting but harder).

\begin{centering}
\begin{figure*}
\includegraphics[width=0.49 \textwidth]{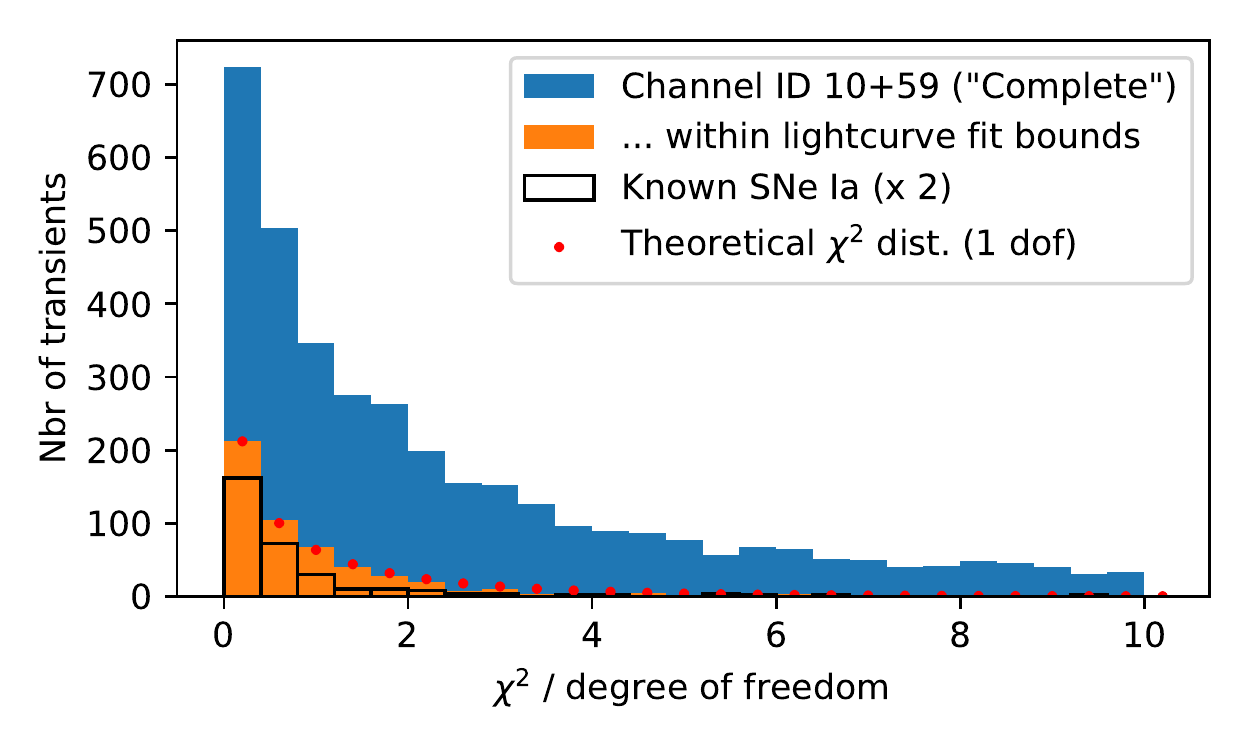}
\includegraphics[width=0.49 \textwidth]{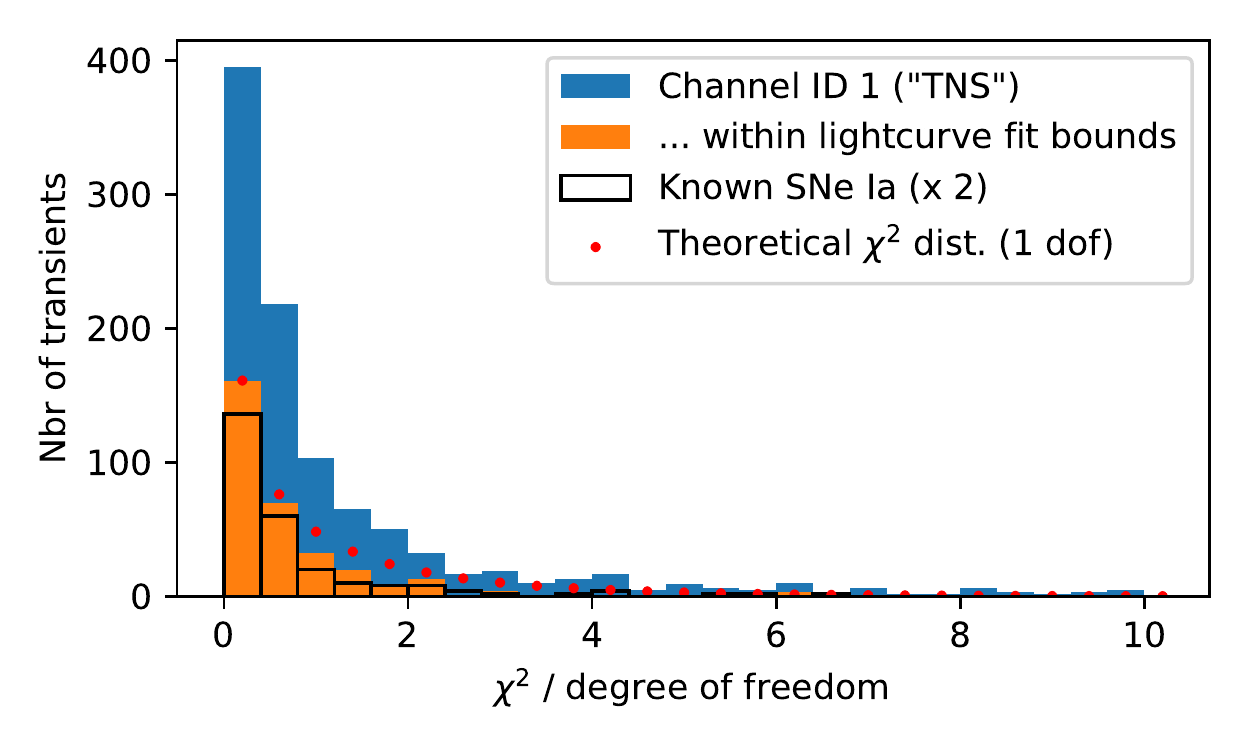}
\caption{Histogram of SALT2 SN Ia fit quality ($chi^2$ per degree of freedom) for the complete $10+59$ channel. Blue bars show the full sample (with enough detections for fit) while orange shows the subset which also fulfill the expected fit parameter requirements. These are compared with with fit quality for the subset of known SN Ia in the comparison sample (outlined bars, scaled with a factor $2$) as well as a standard $\chi^2$ distribution for one degree of freedom (scaled to match the first bin of the restricted sample ). 
}
  \label{fig:chidist}
\end{figure*}
\end{centering}

Out of the $11112$ transients accepted by the complete ($10+59$) channel, $6995$ have the minimal number of detections ($5$) required to fit the SALT2 parameters $x_1$ (lightcurve width), $c$ (lightcurve color), $t_0$ (time of peak light), $x_0$ (peak magnitude) and $z_{phot}$ (redshift from template fit). Further requiring the central values of the fit parameters to match parameter ranges observed among nearby SNe Ia ($-3<x_1<3$, $-1<c<2$, $0.001 < z_{phot}<0.2$ and $z_{err}<0.1$) leaves $634$ transients. In fig.~\ref{fig:chidist}  we compare the distributions of $\chi^2$ per degree of freedom for these samples. We find that the subset following typical SN Ia parameters match both the expected theoretical fit quality distribution and has a distribution similar to the values obtained for the comparison sample of spectroscopically confirmed SNe Ia.
This ``SN Ia compatible'' subset can be thus be used as an approximate photometric SN Ia sample\footnote{Any algorithm for evaluating photometric data can similarly be implemented as a T2 unit and applied to the same rerun dataset. Transient models that can be incorporated into \texttt{SNCOSMO} can even use the same T2 unit and only vary run configuration. 
}.
Repeating this study for the ``TNS'' channel $1$, which accepted $2968$ transients, we find that $1342$ objects can be fit and that out of these $349$ are compatible with the standard SN Ia parameter expectations.

\begin{figure*}
\includegraphics[width=0.49 \textwidth]{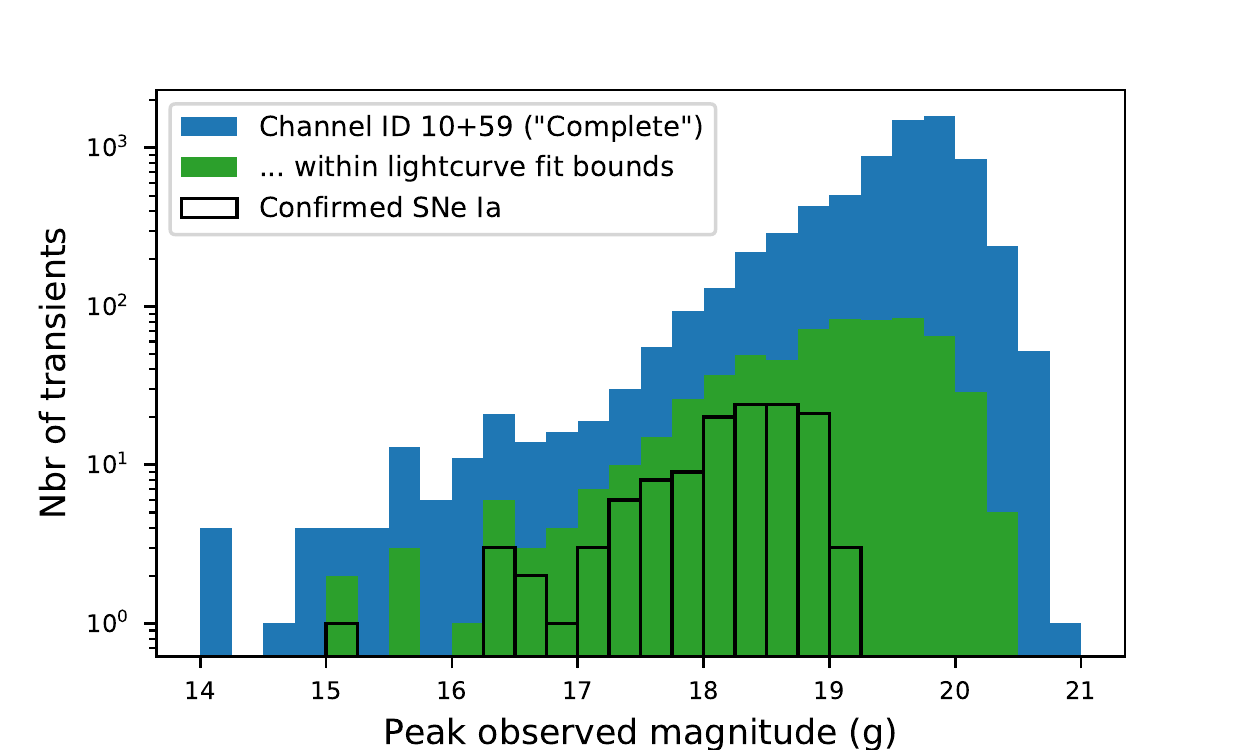}
\includegraphics[width=0.49 \textwidth]{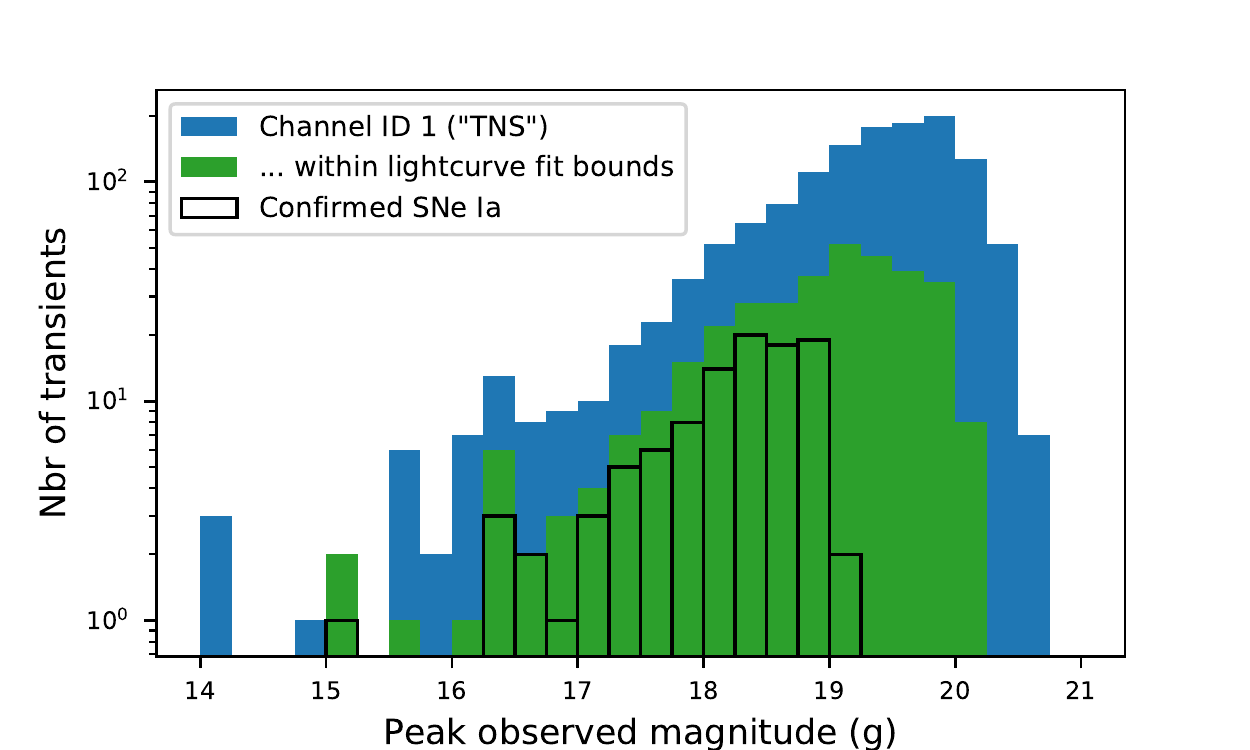}
\caption{ Peak magnitude distributions (ZTF g band) for the same subsets. The comparison sample is not scaled.
  \emph{Left panel:} Data for the complete $10+59$ channel. 
  \emph{Right panels:} Data for the efficient $1$ channel. 
}
  \label{fig:magdist}
\end{figure*}

We next examine the observed peak magnitudes for both the complete and efficient channels (fig~\ref{fig:magdist}). For both channels, the subsets restricted to standard SN Ia parameter ranges agree well with the comparison objects for bright magnitudes ($ < 18.5$ mag).
Fainter than this limit both channels contain a large sample of likely SNe Ia with a  detection efficiency that rapidly drops beyond $19.5$ mag. Both limits are expected as the ZTF RCF program attempts to classify all extragalactic transients brighter than $18.5$ mag and supernovae peaking fainter than $\sim 19.5$ mag often do not yield the five \emph{significant} measurements that are required to trigger the production of an alert and will thus not be included in the lightcurve fit. Most of these fainter SNe will have several late-time observations below the $5 \sigma$ threshold that did not trigger alerts but which will be recoverable once the ZTF image data is released.
We find no significant differences between the complete and TNS channels in terms of magnitude coverage, consistent with the fact that they differ mainly in that the complete channel accepts transients closer to Gaia sources.

\todo{What about the good quality fit object that were not part of reference sample? Examine these visually? (not that many)}

We can thus define two (overlapping) subsets for each channel: The comparison sample of known SN Ia (``Reference SN Ia'') and the photometric SNe Ia (``Photo SN Ia'') with lightcurve fit parameters compatible with a SN Ia. We complement these with five subsets based on external properties:
\begin{itemize}
\item Transients that coincide with an AGN in the Million Quasar Catalog or SDSS QSO catalogs  are marked as ``Known AGN''.
  \item Transients that coincide with the core of a photometric SDSS galaxy are marked ``SDSS core'' (distance less than $1$\arcsec).
  \item Transients that coincide with a SDSS galaxy outside the core are marked ``SDSS off-core'' (distance larger than $1$\arcsec).
  \item Transients that were reported to the TNS as a likely extragalactic transient but do not have a confirmed classification are marked ``TNS AT''.
  \item Transients that do have a TNS classification but are not part of the reference sample of SNe Ia are called ``TNS SN (other)''
\end{itemize}

\begin{figure*}
\includegraphics[width=0.32 \textwidth,clip,trim=1.cm 0 3.5cm 0]{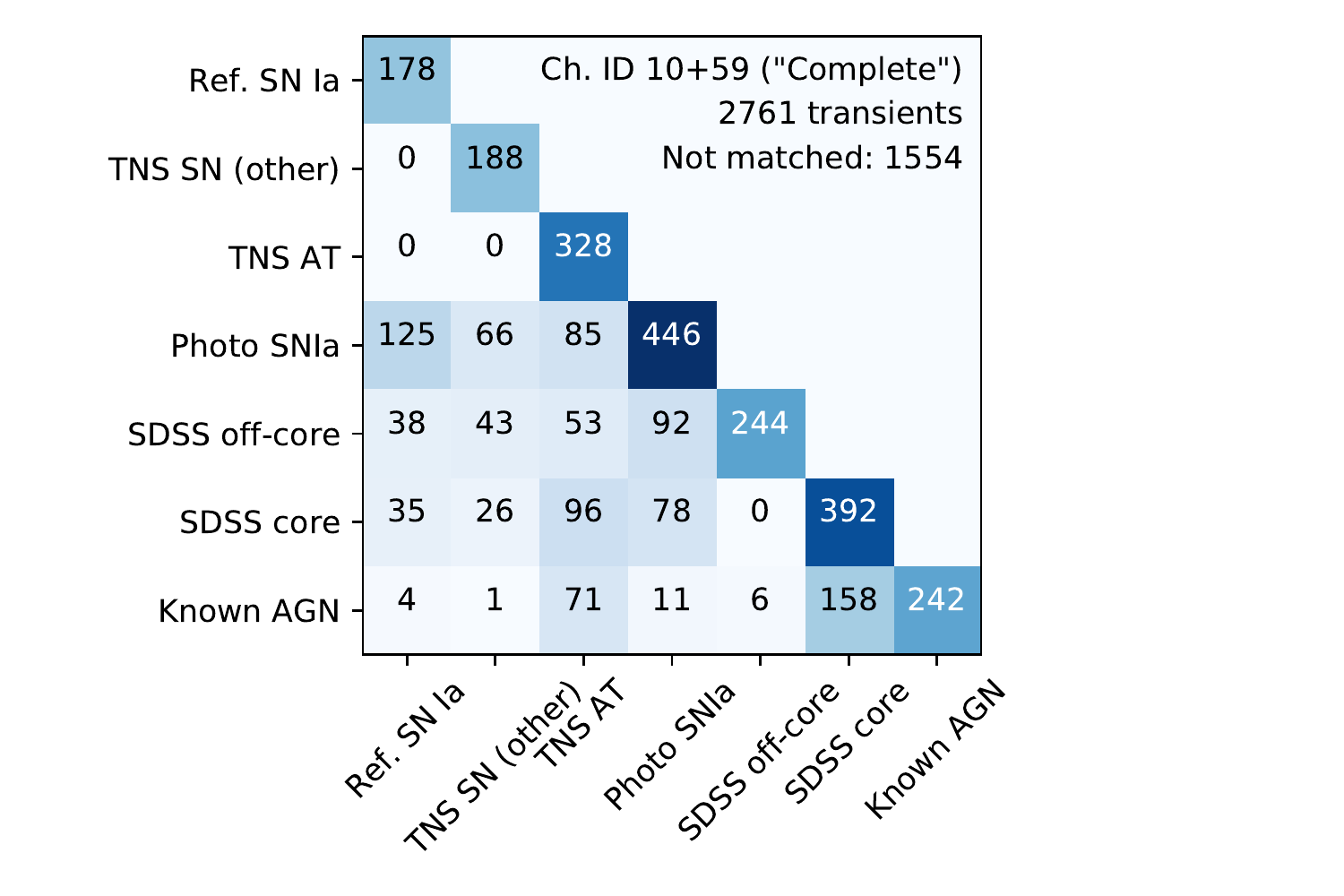}
\includegraphics[width=0.32 \textwidth,clip,trim=1.cm 0 3.5cm 0]{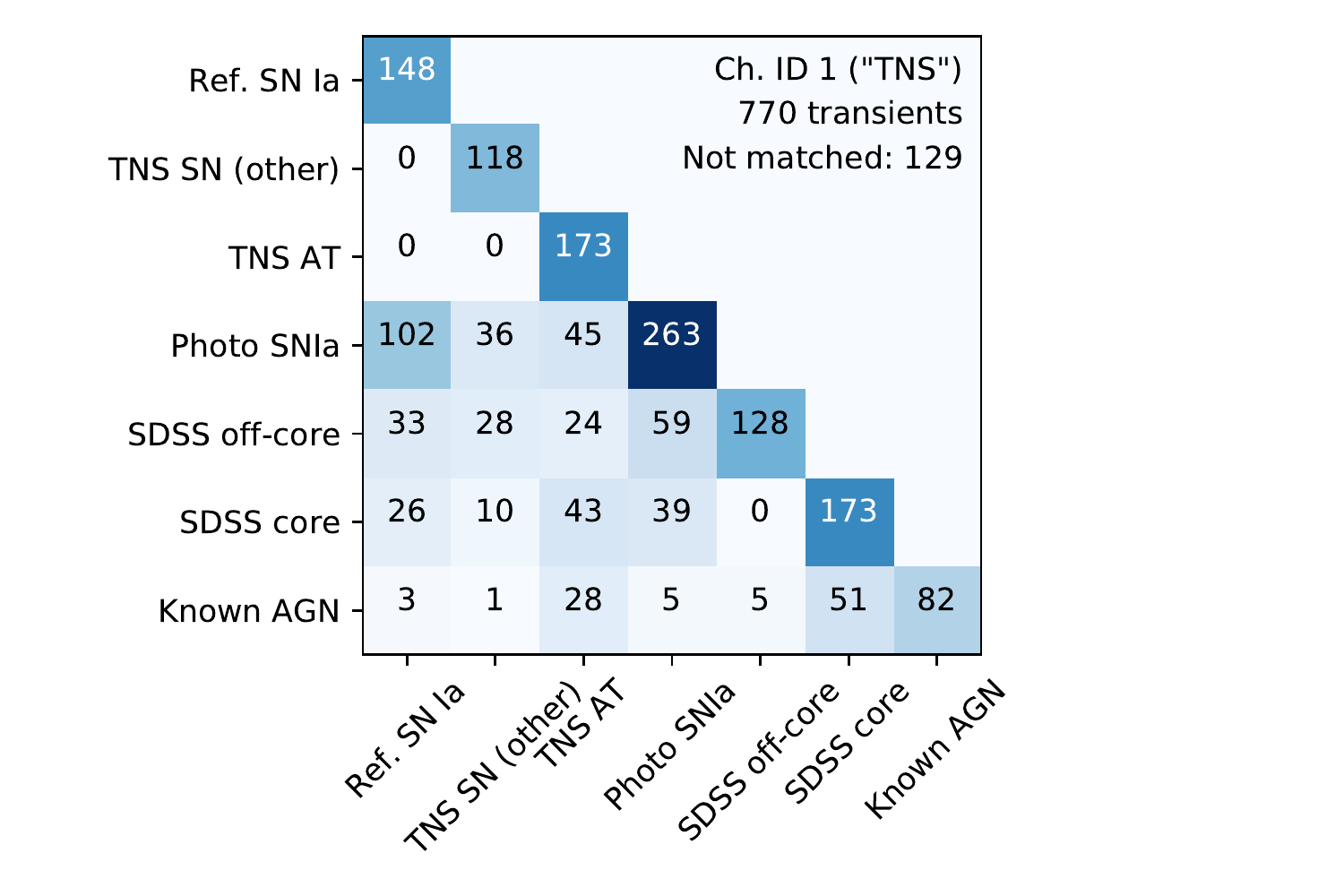}
\includegraphics[width=0.32 \textwidth,clip,trim=1.cm 0 3.5cm 0]{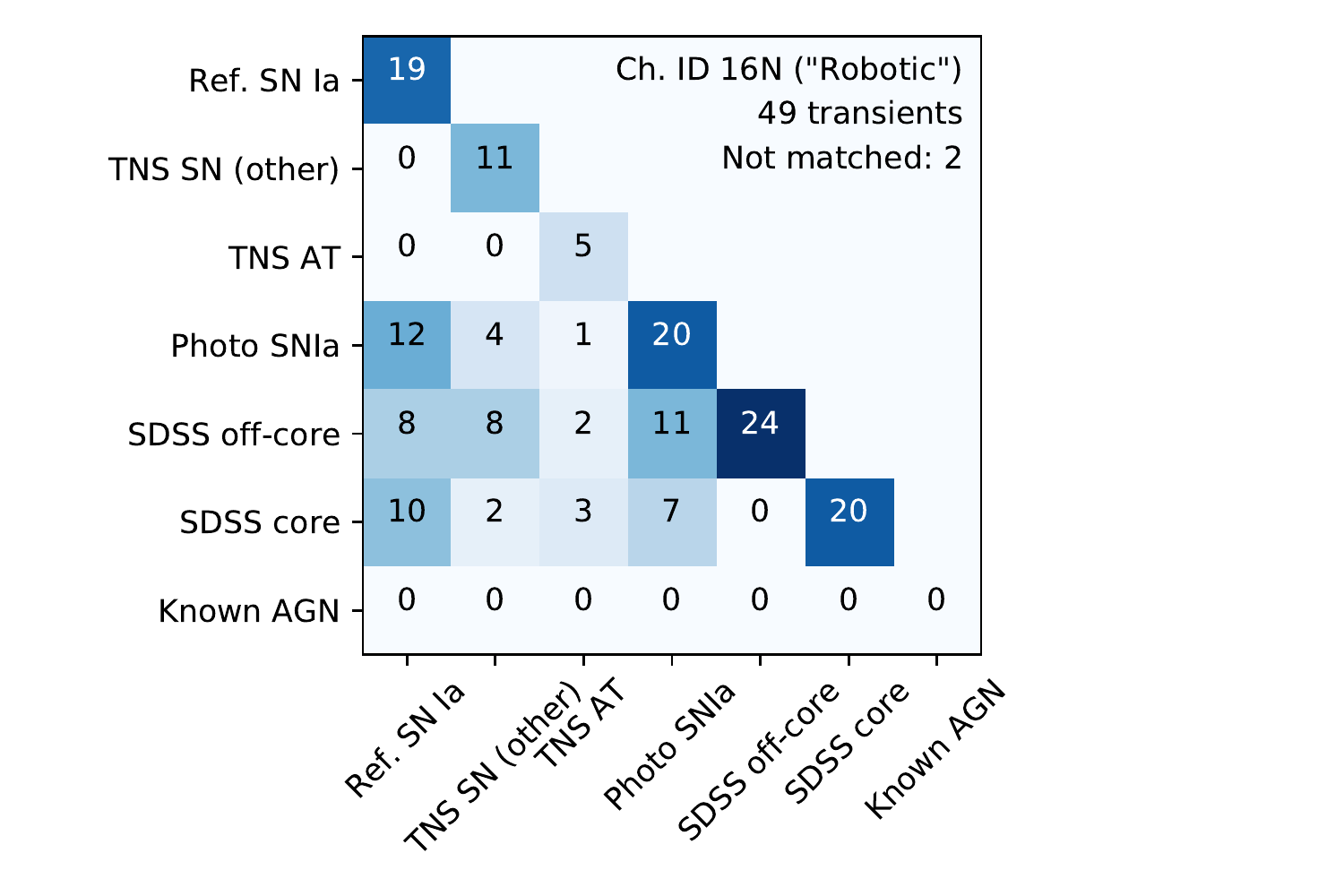}
\caption{Estimated transient types for objects with a peak magnitude brighter than $19.5$ for the channels $10+59$ (``complete''), $1$ (``efficient'') and $16$ (``robotic''). The channel $16$ selection also requires transients to be close to host galaxies with a spectroscopic $z<0.1$ and not in any registered AGN galaxy. 
}
  \label{fig:matrix}
\end{figure*}

The count and overlap between these groups are shown in fig.~\ref{fig:matrix}. We here only include transients with a peak brighter than $19.5$ mag as the fraction with lightcurve fit falls quickly beyond this limit (fig.~\ref{fig:magdist}). We can make several observations already based on this crude accounting:
For the complete channel these categorizations accounts for $40\%$ of all accepted transients. The remaining fraction consists of a combination of real extragalactic transients that were not reported to the TNS, stellar variables not listed in Gaia DR2 and ``noise''.
For the efficient channel, only $20\%$ of all detections ($152$ of $771$) are in this sense unaccounted for.
We observe that large fractions of SNe are found both aligned with the core of SDSS galaxies as well as without association to a photometric SDSS galaxy. This directly demonstrates how care must be taken when selecting targets for surveys looking for complete samples.

A main goal for transient astronomy, and \am,  during the coming decade will be to decrease the fraction of unknown transients as much as possible. Machine learning based photometric classification will be essential to this endeavor, but other developments are as critical. These include the possibility to better distinguish image and subtraction noise (``bogus'') and the ability to compare with calibrated catalogs containing previous variability history.
We plan to revisit this question once the ZTF data can be investigated for previous or later detections.

\subsection{Real-time matching with IceCube neutrino detections}\label{sec:neutrino}

The capabilities and flexibility of \am can also be highlighted through the example of the IceCube realtime neutrino multi-messenger program. Several years ago, the IceCube Neutrino Observatory discovered a diffuse flux of high-energy astrophysical neutrinos \citep{2013Sci...342E...1I}.
Despite recent evidence identifying a flaring blazar as the first neutrino source \citep{2018Sci...361.1378I}, the origin of the bulk of the observed diffuse neutrino flux remains, as yet, undiscovered.
One promising approach to identify these neutrino sources is through multi-messenger programs which explore the possibility of detecting multi-wavelength counterparts to detected neutrinos. Likely high-energy neutrino source classes with optical counterpart are typically variables or transients emitting on timescales of hours to months, for example core collapse supernovae, active galactic nuclei or tidal disruption events \citep{1995PhRvL..75..386W, 2001PhRvL..87v1102A, 2003ApJ...586...79A, 2009ApJ...693..329F, 2013PhRvL.111l1102M, 2015MNRAS.448.2412P, 2016PhRvD..93h3003S, 2017PhRvD..95l3001L, 2017ApJ...838....3S, 2017MNRAS.469.1354D}.  To detect counterparts on these timescales, telescopes are required which feature a high cadence and a large field-of-view, in order to cover a significant fraction of the sky. In addition to an optimized volumetric survey speed capable of discovering large numbers of objects, neutrino correlation studies require robustly-classified samples of optical transient populations. In order to provide a prompt response to selected events within large data volumes, a software framework is required that can analyze and combine optical data streams with real-time multi-messenger data streams. 

\new{Two complementary strategies to search for optical transients in the vicinity of the neutrino sources are currently active in \am.
  Firstly, a target-of-opportunity T0 filter selects ZTF alerts which pass image quality cuts while being spatially and temporally coincident with public IceCube High-Energy neutrino alerts distributed via GCN notifications. This enables rapid follow-up of potentially interesting counterparts, but is only feasible for the handful of neutrinos which have sufficiently large energy to identify them as having a likely astrophysical origin.
  A second program therefore seeks to exploit the more numerous collection of lower-energy astrophysical neutrinos detected by IceCube, which are hidden among a much larger sample of atmospheric background neutrinos. We therefore created a T2 module which in real-time performs a maximum likelihood calculation of the correlation between incoming alerts and an external database of recent neutrino detections. This calculation is based on both spatial and temporal coincidence as well as the estimated neutrino energy. In particular, the consistency of the lightcurve with a given transient class, and the consistency of the neutrino arrival times with the emission models expected for that class, enable us to greatly reduce the number of chance coincidences between neutrinos and optical transients.
  The IceCube collaboration is currently using this setup to search for individual neutrinos or neutrino clusters likely to have an astrophysical origin but with too low energy to warrant an individual GCN notice.
  \nnew{The neutrino DB is populated by the IceCube collaboration in real-time with O(100) neutrinos per day} with directional, temporal and energy information \citep{2017APh....92...30A}. Output is provided as a daily summary of potential matches sent to the IceCube Slack.
  This program allows a systematic selection of which transients to subsequently follow-up spectroscopically. The final sample will provide a magnitude-limited complete, typed catalog of all optical transients which are coincident with neutrinos which can be used to probe neutrino emission from a source population.
}
  
\section{Discussion}\label{sec:disc}

\subsection{The \am TNS stream for new, extragalactic transients}\label{sec:tns}

Most astronomers looking for extragalactic transients have similar requests: A candidate feed which is made available as fast as possible with a large fraction of young supernovae and/or AGNs. By definition, young candidates will not have a lot of detections and the potential gain from photometric classifiers is limited. The efficient TNS channel defined above fulfills these criteria as a large fraction of the comparison sample is recovered while the overall channel count is manageable. Most confirmed SNe Ia were detected more than $10$ days before peak, confirming the potential for early detections.

To allow the community fast access to these transients, we use channel ID$1$ (``TNS'') to automatically submit all ZTF detections from the MSIP program as public astronomical transients to the TNS using senders starting with the identifier \texttt{ZTF\_AMPEL}. AT 2019abn (ZTF19aadyppr) in the Messier 51 (Whirlpool Galaxy) provides an example of this process. AT 2019abn was observed by ZTF at JD $2458509.0076$ and reported to the TNS by \am slightly more than one hour later.

\begin{figure}
  \begin{centering}
\includegraphics[width=0.49 \textwidth]{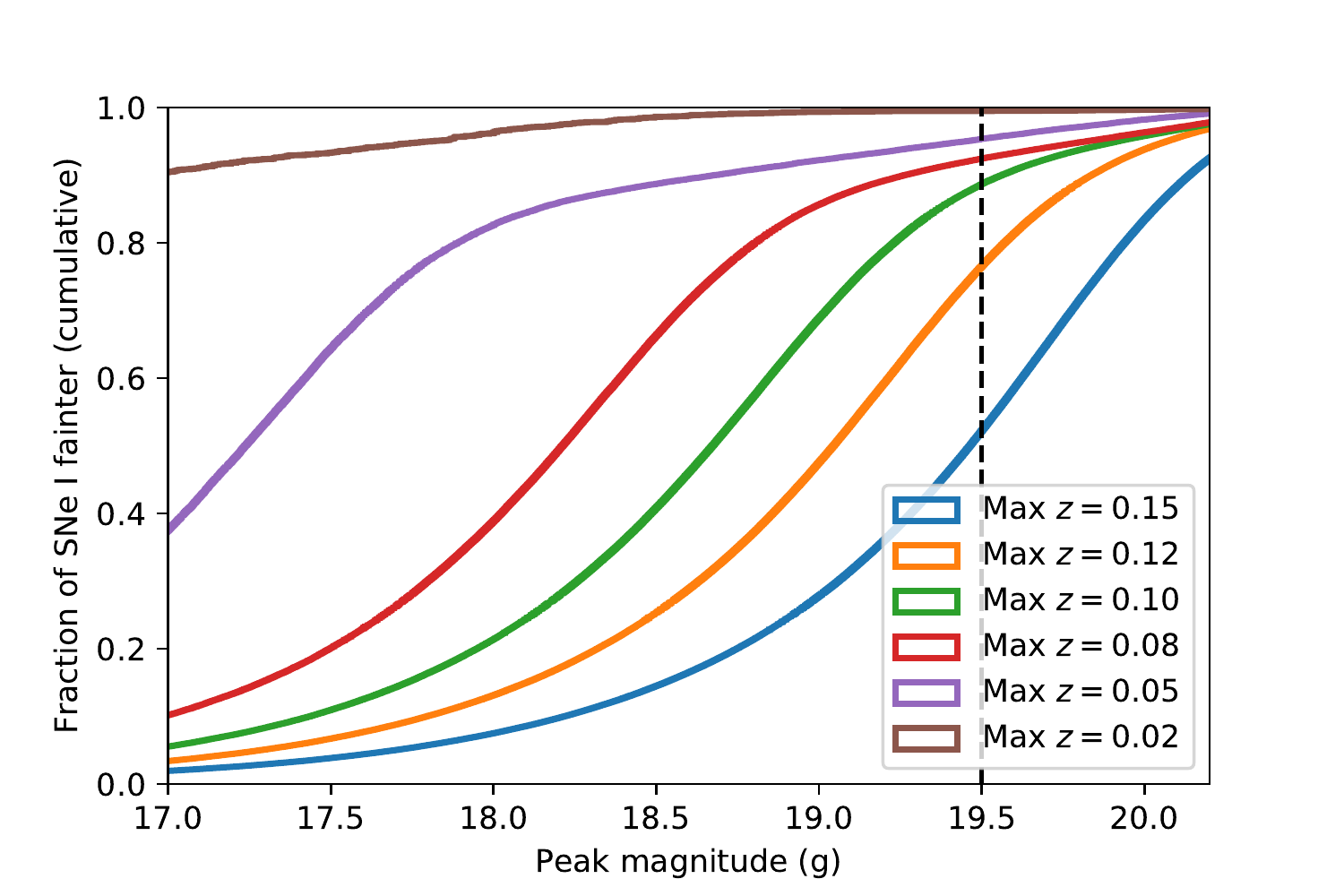}
\caption{Cumulative \texttt{simsurvey} peak magnitude for simulated data, divided according to max redshift. Dashed lines show the current $19.5$ depth of \am TNS submissions.
}
\label{fig:magcum}
\end{centering}
\end{figure}

To make the published candidate stream even more pure, the following additional cuts are made prior to submission. First, we restrict the sample to transients brighter than $19.5$ mag (the limit to which the channel content study was carried out).  The magnitude depth will be increased once a sufficiently-low stellar contamination rate has been confirmed for fainter transients. Fig.~\ref{fig:magcum} shows the expected cumulative distributions of peak magnitudes for SNe Ia below different redshift limits as determined by \texttt{simsurvey}. A $19.5$ mag peak limit implies a $\sim 90\%$ completeness for SNe Ia at $z<0.08$ based on the expected magnitude distribution. For the volumetric completeness this should be combined with the $80\%$ coverage completeness determined above (which is mainly driven by sky position). We currently only submit candidates found above a galactic latitude of $14$ degrees to reduce contamination by stellar variables. An inspection of the so far reported candidates find less than $5\%$ to be of likely stellar origin.
Candidates compatible with known AGN/QSOs are marked as such in the TNS comment field. \rep{TNS users looking for the purest SN stream can thus disregard any transients with this comment.} 

Two TNS bots are currently active: \texttt{ZTF\_AMPEL\_NEW} specifically aims to submit only young candidates with a significant non-detection available within $5$ days prior to detection and no history of previous variability. This will create a bias against AGNs with repeated, isolated variability as well as transients with a long, slow rise-time but further rejects variable stars and provides a quick way to find follow-up targets.
A second sender, \texttt{ZTF\_AMPEL\_COMPLETE}, only requires a non-detection within the previous $30$ days.\footnote{These bots replace the initial \texttt{ZTF\_AMPEL\_MSIP} sender, which is no longer in use.}

In summary, \rep{the live submission of \am detections to the TNS provides} a high quality feed for anyone looking for new, extragalactic transients brighter than $19.5$ mag. The contamination by variable stars is estimated at $<5\%$, the fraction of SNe to be $>50 \%$ and for SNe Ia with a peak brighter than $\sim 18.5$ mag the SN Ia completeness is $80\%$, out of which $\sim 60\%$ will be detected prior to ten days before lightcurve peak.
\todo{From simulation? WRITE WHAT FRACTION PEAKED ABOVE MAG 19.5, INDEPENDENT OF NBR OF DETECTIONS, WHICH IS WHAT WOULD BE SUBMITTED.} Extrapolating rates from the four (summer) month ZTF rerun would predict this program to submit $\sim 9000$ astronomical transients to the TNS each year. The breaks due to typical Palomar winter weather makes this an upper limit.

\subsection{\rep{Work towards an \am testing and rerun environment}}

\rep{The next \am version is already being developed. We plan for this to contain an interface where users can directly upload channel and unit configurations and have them process a stream of archived alerts. The container generation means that such a configuration could be automatically spun up in an automatic and secure mode at a computer center. This run environment would allow both more complete tests as well as more flexibility in carrying out large scale reruns.

  }

\section{Conclusions}

We here introduce \am as a comprehensive tool for working with streams of astronomical data.
More and more facilities provide real-time data shaped into streams, which creates opportunities to make new discoveries while emphasizing the challenge in that also actions \emph{not} taken are scientific choices.  \am includes tools for brokering (distributing), analyzing, selecting and reacting to transients. 
Users contribute channels, which regulate how transients are processed at four internal tiers. The implementation was guided by our suggestions for how to face these new opportunities and challenges for transient analysis:
\begin{itemize}
\item \emph{Provenance and reproducibility} are guaranteed by the combination of information stored in a permanent database, containerized software and alert archive in a system designed to allow autonomous analysis chains.
\item A modular system provides \emph{analysis flexibility}, and introduces a method for developers to allow software distribution and referencing.
\item The combination of these two capabilities allows users to track the impact of \emph{versions} of both data and software.
  \item Finally, the database has been designed to manage the \emph{alert rates} expected from surveys such as LSST. 
\end{itemize}
A requirement for achieving this was the division of the alert processing into four tiers and the recognition that each transient is connected to a growing set of \emph{states}, each of which consists of a specified set of \emph{datapoints}. A \emph{transient view} collects the information of a transient available at a given time.

Three sample uses of \am were introduced. 
We first used a reprocessing of alerts from the first four months of ZTF operations to create a ``cooking book'' of filter definitions with defined acceptance and completeness rates. As part of this study, we show that ZTF detected and issued alerts for all SNe Ia reported to the TNS, and that \am can operate at the high data rates expected for LSST. \new{Three channels were highlighted: A ``complete'' channel recovering all known SN Ia with a comparably small total count, a ``TNS'' channel which allows SNe to be detected early and efficiently, and a small ``robotic'' channel which can serve as a starting point for automatic follow-up observations. Channel/program distinctions along these lines will become natural as astronomers tap into future large transient flows.}
We then took a first step in identifying the content of these three channels. For the complete channel the fraction of real extragalactic transients is estimated to be larger than $40\%$ and for the TNS channel above $80\%$. The robotic channel is designed to retain only target transients in known nearby galaxies. We plan to continue reprocessing alerts with refined analysis units, improved photometry and larger alert sets.
As a third example, we introduce the live correlation analysis between optical ZTF alerts and candidate extragalactic neutrinos from IceCube, where a T2 unit calculates test statistics between all potential matches and selects targets for spectroscopic follow-up. This methodology can be directly applied to other kinds of multi-messenger studies.

The \am live instance processes the ZTF alert stream and anyone can become a user through creating a channel following the guidelines available at the \texttt{AmpelProject/Ampel-contrib-sample} github repository. However, as many astronomers are interested in similar objects, \am also provides  a more immediate avenue to likely young, extragalactic transients through a real-time propagation of high-quality candidates to the TNS. The chosen channel configuration (``TNS'', ID$1$) was shown to detect $\sim 80\%$ the SNe in the comparison sample, with more than $50\%$ detected prior to phase $-10$ days. This setup is expected to provide O(1000) astronomical transients each year.

\todo{discuss more that if input data is a stream the whole analysis has to happen in a streaming context.}

\begin{acknowledgements}

Based on observations obtained with the Samuel Oschin Telescope 48-inch and the 60-inch Telescope at the Palomar Observatory as part of the Zwicky Transient Facility project. ZTF is supported by the National Science Foundation under Grant No. AST-1440341 and a collaboration including Caltech, IPAC, the Weizmann Institute for Science, the Oskar Klein Center at Stockholm University, the University of Maryland, the University of Washington, Deutsches Elektronen-Synchrotron and Humboldt University, Los Alamos National Laboratories, the TANGO Consortium of Taiwan, the University of Wisconsin at Milwaukee, and Lawrence Berkeley National Laboratories. Operations are conducted by COO, IPAC, and UW. \nnew{The authors are grateful to the IceCube Collaboration for providing the neutrino dataset and supporting its use with \am}.
N. M. acknowledges the support of the Helmholtz Einstein International Berlin Research School in Data Science (HEIBRiDS), Deutsches Elektronensynchrotron (DESY), and Humboldt-Universität zu Berlin.
M. R. acknowledges the support from the European Research Council (ERC) under the European Union’s Horizon 2020 research and innovation program (grant agreement no 759194 - USNAC).

\end{acknowledgements}

\bibliographystyle{aa} 
\bibliography{ampel.bib} 

\appendix

\section{Creating the TNS comparison sample}\label{app:matchlist}

The comparison sample that is used to estimate channel efficiencies was constructed through retrieving all TNS SNe classified as Type Ia supernovae (not including peculiar subtypes) and with a detection date between June 5th and September 15h 2018. This was further restricted to SNe above an absolute galactic latitude of $14$ degrees. This leaves $310$ objects shown in Table~\ref{tab:matchlist}. Out of these $20$ has positions outside the ZTF MSIP primary field grid and $8$ were projected to land in gaps between ZTF CCDs or within the $1$ \% of chip pixels closes to a readout edge.

As ZTF field references have been continuously produced during the first season of operations we also verify that subtractions were made at least $3$ days prior to the SN detection. For $89$ SNe no references were available while for $58$ SNe a reference was only available in either $g$ or $r$ band. One TNS object included in this list, SN 2018ekt, was as part of this study found to have been erroneously classified (this has thus now been removed). Excluding this leaves a main comparison sample of $134$ SNe Ia that were observed by ZTF in the nominal ZTF MSIP cadence. 
Among SNe only found in one band SN 2018fvh is located on bad pixels, SN 2018cmu is only detected in a single alert (one detection) and SN 2018cmk was detected by ZTF but more than $3$ arcsec from the reported TNS position.\todo{JS: 8 out of 240 on chip gap, 3 percent ? Answer with buffer yes.}

\clearpage
\onecolumn

\begin{center}
{\tiny
\begin{longtable}{|c|c|c|c|c|c|c|c|}
\caption{TNS SNe Ia detected between June 5th and Sep 15th.}\\
\hline
{Name} & {RA} & {DEC} & {Redshift} & {ZTF field match} & {Detectable} \\
\hline
\endfirsthead
\hline
{Name} & {RA} & {DEC} & {Redshift} & {ZTF field match} & {Detectable} \\
\hline
\endhead
\hline
\endfoot
\hline
\endlastfoot
\input{searchtable}
\label{tab:matchlist}
\end{longtable}
}
\end{center}

\clearpage
\twocolumn

\end{document}

%% file: channeltableGeneral.tex
11 & 0.5 & 4 & multinight & good & nominal & 125 & 1857 & 0.956 & 0.067 & 120 & 1514 & 0.953 & 0.047 \\
26 & 0.3 & 4 & new & excellent & nominal & 36 & 293 & 1.000 & 0.296 & 35 & 257 & 1.000 & 0.308 \\
34 & 0.5 & 2 & multinight & excellent & nominal & 128 & 2479 & 0.944 & 0.079 & 123 & 1964 & 0.941 & 0.059 \\
18 & 0.5 & 6 & new & all & nominal & 2 & 34 & 0.000 & 0.000 & 2 & 28 & 0.000 & 0.000 \\
77 & 0.3 & 4 & persistent & all & moderate & 131 & 10672 & 0.832 & 0.000 & 126 & 9705 & 0.824 & 0.000 \\
51 & 0.3 & 4 & new & good & nominal & 39 & 357 & 1.000 & 0.321 & 37 & 311 & 1.000 & 0.308 \\
57 & 0.3 & 4 & persistent & good & nominal & 130 & 3078 & 0.830 & 0.000 & 125 & 2351 & 0.822 & 0.000 \\
64 & 0.3 & 8 & multinight & good & nominal & 76 & 580 & 0.554 & 0.000 & 74 & 524 & 0.556 & 0.000 \\
1 & 0.3 & 2 & new & good & nominal & 110 & 2968 & 0.987 & 0.557 & 106 & 2562 & 0.987 & 0.533 \\
28 & 0.5 & 2 & new & excellent & nominal & 95 & 1833 & 0.985 & 0.485 & 92 & 1547 & 0.985 & 0.462 \\
4 & 0.5 & 2 & new & good & nominal & 103 & 2286 & 0.986 & 0.514 & 99 & 1909 & 0.985 & 0.485 \\
10 & 0.5 & 2 & multinight & good & nominal & & & & & & & &  \\
59 & 0.3 & 4 & persistent & good & moderate & & & & &  & &  &  \\
10+59 & & & &  & & 134 & 11112 & 0.926 & 0.084 & 129 & 9973 & 0.923 & 0.055 \\

%% file: channeltableNearby.tex
85 & 0.3 & 4 & persistent & good & nominal & 21 & 261 & 0.750 & 0.000 & 19 & 143 & 0.714 & 0.000 \\
34 & 0.5 & 2 & multinight & excellent & nominal & 20 & 231 & 0.867 & 0.067 & 18 & 136 & 0.846 & 0.077 \\
16 & 0.5 & 2 & new & all & nominal & 17 & 160 & 0.923 & 0.385 & 15 & 101 & 0.909 & 0.273 \\
29 & 0.5 & 4 & new & excellent & nominal & 4 & 9 & 1.000 & 0.000 & 3 & 6 & 1.000 & 0.000 \\
84 & 0.3 & 8 & multinight & all & moderate & 11 & 32 & 0.333 & 0.000 & 10 & 24 & 0.375 & 0.000 \\
28 & 0.5 & 2 & new & excellent & nominal & 15 & 117 & 0.917 & 0.333 & 14 & 79 & 0.909 & 0.273 \\

%% file: searchtable.tex
SN 2018guh & 21:53:36.95 & +07:14:06.30 & 0.120  & In MSIP grid &  Detectable g or r \\
SN 2018guc & 02:30:57.89 & +13:26:22.02 & 0.091  & In MSIP grid &  Detectable g or r \\
SN 2018gtg & 01:57:26.22 & +08:55:02.54 & 0.108  & In MSIP grid &  Detectable g+r \\
SN 2018gsb & 16:46:08.75 & +19:27:17.22 & 0.060  & In MSIP grid &  Detectable g+r \\
SN 2018gqo & 08:27:18.13 & +10:59:57.28 & 0.087  & In MSIP grid &  Detectable g or r \\
SN 2018gqn & 08:42:44.87 & +10:41:42.56 & 0.060  & In MSIP grid &  Detectable g or r \\
SN 2018gqa & 17:51:11.31 & +53:18:52.09 & 0.105  & In MSIP grid &  Detectable g+r \\
SN 2018gkz & 07:58:11.54 & +19:31:07.92 & 0.240  & In MSIP grid &  Detectable g or r \\
SN 2018gkw & 07:57:37.92 & +44:44:48.60 & 0.072  & In MSIP grid &  Detectable g or r \\
SN 2018gkj & 22:02:27.07 & -64:21:06.77 & 0.042  & In MSIP grid &  No reference \\
SN 2018gka & 02:59:20.01 & -14:05:59.27 & 0.100  & In MSIP grid &  No reference \\
SN 2018gjz & 20:51:09.99 & -24:53:38.86 & 0.083  & Chip / RC border & Not clear \\
SN 2018git & 01:40:04.26 & +15:04:16.67 & 0.071  & In MSIP grid &  Detectable g+r \\
SN 2018gif & 03:27:15.75 & +08:42:39.13 & 0.039  & In MSIP grid &  No reference \\
SN 2018ghr & 22:59:12.79 & +00:24:17.96 & 0.072  & In MSIP grid &  Detectable g or r \\
SN 2018ghp & 00:13:42.06 & +27:34:03.45 & 0.069  & In MSIP grid &  Detectable g or r \\
SN 2018ghh & 10:14:20.24 & +56:19:16.44 & 0.050  & In MSIP grid &  Detectable g+r \\
SN 2018ggz & 01:47:53.66 & +42:11:45.49 & 0.074  & In MSIP grid &  Detectable g or r \\
SN 2018ggx & 01:04:38.80 & -04:15:29.42 & 0.038  & In MSIP grid &  Detectable g or r \\
SN 2018ggw & 18:28:15.30 & +46:21:12.27 & 0.053  & In MSIP grid &  Detectable g+r \\
SN 2018ggt & 00:57:02.61 & -00:52:25.68 & 0.044  & In MSIP grid &  Detectable g+r \\
SN 2018ggq & 18:37:17.16 & +49:08:54.31 & 0.090  & In MSIP grid &  Detectable g+r \\
SN 2018ggc & 08:38:10.69 & +24:53:26.72 & 0.030  & In MSIP grid &  Detectable g or r \\
SN 2018gfw & 20:30:53.58 & -26:31:04.42 & 0.032  & In MSIP grid &  No reference \\
SN 2018gfs & 15:52:03.30 & +30:44:05.71 & 0.080  & In MSIP grid &  Detectable g+r \\
SN 2018gfi & 00:58:53.95 & -24:11:47.22 & 0.020  & Outside MSIP grid & Not observable \\
SN 2018gfg & 20:59:09.74 & -14:15:47.47 & 0.082  & In MSIP grid &  Detectable g or r \\
SN 2018gff & 02:05:40.56 & +19:05:31.82 & 0.069  & In MSIP grid &  Detectable g or r \\
SN 2018gfe & 16:17:48.37 & +41:28:10.37 & 0.065  & In MSIP grid &  Detectable g+r \\
SN 2018ges & 21:33:50.41 & +24:20:03.41 & 0.115  & Chip / RC border & Not clear \\
SN 2018geo & 16:18:13.84 & +39:07:25.74 & 0.031  & In MSIP grid &  Detectable g+r \\
SN 2018gec & 07:09:53.50 & +54:58:07.46 & 0.037  & Outside MSIP grid & Not observable \\
SN 2018gdg & 00:30:24.63 & -01:15:50.88 & 0.057  & In MSIP grid &  Detectable g+r \\
SN 2018gck & 00:50:56.61 & +03:29:55.02 & 0.100  & In MSIP grid &  Detectable g or r \\
SN 2018gcg & 17:16:10.37 & +01:25:46.25 & 0.094  & In MSIP grid &  Detectable g+r \\
SN 2018fzw & 05:42:48.04 & -24:29:34.39 & 0.046  & In MSIP grid &  No reference \\
SN 2018fzm & 21:03:19.79 & -51:32:48.23 & 0.047  & In MSIP grid &  No reference \\
SN 2018fzi & 23:39:47.31 & -23:55:12.81 & 0.080  & In MSIP grid &  Detectable g or r \\
SN 2018fyg & 02:45:59.36 & +42:32:29.67 & 0.039  & In MSIP grid &  No reference \\
SN 2018fwn & 03:29:34.39 & -19:41:41.35 & 0.044  & In MSIP grid &  No reference \\
SN 2018fwi & 22:47:46.56 & -31:15:33.70 & 0.115  & In MSIP grid &  No reference \\
SN 2018fvy & 01:32:16.23 & -33:06:05.61 & 0.036  & In MSIP grid &  No reference \\
SN 2018fvw & 21:02:20.23 & +19:57:46.25 & 0.040  & In MSIP grid &  Detectable g+r \\
SN 2018fvr & 15:52:22.18 & +22:55:56.41 & 0.053  & In MSIP grid &  Detectable g+r \\
SN 2018fvq & 21:31:32.38 & -27:15:43.56 & 0.073  & In MSIP grid &  No reference \\
SN 2018fvm & 20:37:27.79 & -37:43:16.93 & 0.018  & In MSIP grid &  No reference \\
SN 2018fvl & 21:01:12.26 & +14:29:07.68 & 0.066  & In MSIP grid &  Detectable g+r \\
SN 2018fvj & 04:29:31.07 & -11:01:50.56 & 0.066  & In MSIP grid &  No reference \\
SN 2018fvi & 01:57:42.56 & -67:11:12.80 & 0.040  & In MSIP grid &  No reference \\
SN 2018fvh & 01:15:10.21 & -18:10:59.56 & 0.073  & In MSIP grid &  Detectable g or r \\
SN 2018fvb & 16:00:00.65 & -11:05:27.71 & 0.051  & In MSIP grid &  No reference \\
SN 2018fva & 23:32:34.75 & +08:45:10.38 & 0.074  & In MSIP grid &  No reference \\
SN 2018fuu & 23:24:56.60 & +09:25:52.68 & 0.060  & In MSIP grid &  Detectable g or r \\
SN 2018fum & 01:36:11.29 & +27:16:29.51 & 0.060  & In MSIP grid &  Detectable g or r \\
SN 2018fuk & 05:45:08.16 & -79:23:47.52 & 0.018  & In MSIP grid &  No reference \\
SN 2018fuj & 04:30:58.48 & -26:46:54.10 & 0.051  & Outside MSIP grid & Not observable \\
SN 2018fub & 00:40:30.65 & -50:41:15.14 & 0.029  & In MSIP grid &  No reference \\
SN 2018fty & 02:26:47.30 & -09:04:02.32 & 0.054  & In MSIP grid &  Detectable g or r \\
SN 2018ftu & 01:34:41.55 & -05:38:05.25 & 0.063  & In MSIP grid &  Detectable g or r \\
SN 2018ftt & 17:36:27.05 & +22:29:25.96 & 0.056  & In MSIP grid &  Detectable g+r \\
SN 2018fte & 01:44:08.27 & +07:58:37.46 & 0.062  & In MSIP grid &  Detectable g or r \\
SN 2018ftd & 02:01:16.11 & -01:13:26.26 & 0.062  & In MSIP grid &  Detectable g or r \\
SN 2018ftc & 23:28:03.88 & +09:46:13.30 & 0.050  & Outside MSIP grid & Not observable \\
SN 2018fst & 01:19:32.29 & -18:10:24.46 & 0.100  & In MSIP grid &  Detectable g or r \\
SN 2018fss & 03:06:26.81 & +41:12:49.79 & 0.060  & In MSIP grid &  Detectable g or r \\
SN 2018fsr & 07:49:07.34 & +32:58:50.22 & 0.028  & In MSIP grid &  Detectable g or r \\
SN 2018fsf & 22:40:24.67 & +16:11:26.86 & 0.095  & In MSIP grid &  Detectable g+r \\
SN 2018fse & 17:14:14.51 & +36:27:32.30 & 0.084  & In MSIP grid &  Detectable g+r \\
SN 2018fsa & 22:25:37.11 & -13:04:28.27 & 0.043  & In MSIP grid &  No reference \\
SN 2018frx & 17:23:49.83 & +46:18:05.72 & 0.085  & In MSIP grid &  Detectable g+r \\
SN 2018frv & 20:40:31.43 & -46:34:39.22 & 0.043  & In MSIP grid &  No reference \\
SN 2018frk & 22:46:16.77 & -07:50:13.91 & 0.080  & Outside MSIP grid & Not observable \\
SN 2018fqn & 22:39:38.02 & -15:05:08.66 & 0.052  & Outside MSIP grid & Not observable \\
SN 2018fql & 23:41:16.08 & -28:57:44.74 & 0.049  & In MSIP grid &  No reference \\
SN 2018fqk & 22:33:09.02 & +04:07:47.03 & 0.041  & In MSIP grid &  No reference \\
SN 2018fqj & 14:23:49.00 & +34:06:10.35 & 0.100  & In MSIP grid &  Detectable g+r \\
SN 2018fqe & 15:38:03.03 & +51:21:37.30 & 0.074  & In MSIP grid &  Detectable g+r \\
SN 2018fqd & 20:51:51.48 & -36:50:13.31 & 0.039  & Chip / RC border & Not clear \\
SN 2018fqc & 05:20:45.15 & -23:04:04.67 & 0.085  & In MSIP grid &  No reference \\
SN 2018fpv & 15:08:04.21 & +37:15:05.80 & 0.079  & In MSIP grid &  Detectable g+r \\
SN 2018fpe & 15:01:03.65 & +44:09:23.54 & 0.088  & In MSIP grid &  Detectable g+r \\
SN 2018fop & 01:15:18.11 & -06:51:32.54 & 0.020  & In MSIP grid &  No reference \\
SN 2018fod & 16:13:41.56 & +10:39:30.68 & 0.080  & In MSIP grid &  Detectable g or r \\
SN 2018foc & 02:44:07.12 & +37:31:27.46 & 0.031  & In MSIP grid &  Detectable g or r \\
SN 2018fnq & 20:12:30.00 & -44:06:35.14 & 0.019  & Chip / RC border & Not clear \\
SN 2018fng & 16:07:03.01 & +15:35:44.68 & 0.040  & In MSIP grid &  Detectable g+r \\
SN 2018fnf & 23:40:45.92 & +14:06:01.34 & 0.063  & In MSIP grid &  Detectable g or r \\
SN 2018fne & 18:31:33.45 & +49:48:57.94 & 0.045  & In MSIP grid &  Detectable g+r \\
SN 2018fnd & 16:47:54.59 & +42:58:07.61 & 0.075  & In MSIP grid &  Detectable g+r \\
SN 2018fnc & 18:16:07.46 & +70:01:04.60 & 0.080  & In MSIP grid &  Detectable g+r \\
SN 2018fmu & 19:45:40.96 & +66:30:29.85 & 0.085  & In MSIP grid &  No reference \\
SN 2018fmr & 21:54:00.01 & +08:06:52.87 & 0.070  & In MSIP grid &  Detectable g+r \\
SN 2018fli & 15:58:11.68 & +19:45:54.75 & 0.066  & In MSIP grid &  Detectable g or r \\
SN 2018flg & 16:14:46.99 & +43:18:19.06 & 0.060  & In MSIP grid &  Detectable g+r \\
SN 2018fjx & 22:27:36.81 & +22:53:43.12 & 0.057  & In MSIP grid &  Detectable g or r \\
SN 2018fjv & 16:45:29.83 & +65:14:20.60 & 0.070  & In MSIP grid &  Detectable g+r \\
SN 2018fju & 16:19:44.23 & +50:33:06.72 & 0.056  & In MSIP grid &  Detectable g+r \\
SN 2018fjc & 21:38:53.12 & +31:44:15.09 & 0.042  & In MSIP grid &  Detectable g or r \\
SN 2018fja & 21:59:01.79 & +11:06:13.32 & 0.076  & In MSIP grid &  Detectable g+r \\
SN 2018fiw & 00:15:01.84 & +34:48:40.33 & 0.048  & In MSIP grid &  Detectable g+r \\
SN 2018fiv & 21:54:36.51 & +18:06:56.78 & 0.050  & In MSIP grid &  Detectable g+r \\
SN 2018fio & 15:21:14.35 & +30:38:11.91 & 0.075  & In MSIP grid &  Detectable g+r \\
SN 2018fin & 16:51:00.83 & +25:52:33.51 & 0.060  & In MSIP grid &  Detectable g+r \\
SN 2018fim & 16:37:39.98 & +25:19:15.26 & 0.068  & In MSIP grid &  Detectable g+r \\
SN 2018fhx & 06:24:38.04 & -23:43:58.94 & 0.023  & In MSIP grid &  No reference \\
SN 2018fhw & 04:18:06.27 & -63:36:54.25 & 0.017  & In MSIP grid &  No reference \\
SN 2018fhh & 18:48:19.09 & +30:36:25.25 & 0.100  & In MSIP grid &  Detectable g+r \\
SN 2018fhg & 15:03:45.14 & +61:34:04.41 & 0.066  & In MSIP grid &  Detectable g+r \\
SN 2018fhe & 19:50:48.25 & +64:55:22.29 & 0.068  & In MSIP grid &  Detectable g+r \\
SN 2018fgj & 04:47:32.42 & -02:18:22.50 & 0.031  & In MSIP grid &  No reference \\
SN 2018ffo & 01:05:42.50 & +07:32:55.65 & 0.040  & In MSIP grid &  No reference \\
SN 2018ffn & 00:24:07.84 & -07:52:49.64 & 0.100  & Outside MSIP grid & Not observable \\
SN 2018ffi & 21:40:11.30 & +21:33:30.22 & 0.090  & Chip / RC border & Not clear \\
SN 2018ffb & 22:58:11.52 & -20:17:02.76 & 0.070  & In MSIP grid &  No reference \\
SN 2018fey & 01:22:19.50 & -02:29:48.62 & 0.060  & In MSIP grid &  No reference \\
SN 2018few & 01:04:15.49 & -42:44:28.14 & 0.066  & In MSIP grid &  No reference \\
SN 2018fev & 13:46:54.60 & +32:26:34.30 & 0.110  & In MSIP grid &  Detectable g+r \\
SN 2018fem & 13:01:12.62 & +39:45:23.25 & 0.070  & In MSIP grid &  Detectable g+r \\
SN 2018fel & 16:52:31.24 & +23:23:00.67 & 0.100  & In MSIP grid &  Detectable g+r \\
SN 2018fec & 18:00:35.19 & +61:41:51.76 & 0.075  & In MSIP grid &  Detectable g+r \\
SN 2018feb & 17:10:11.16 & +21:38:56.53 & 0.015  & In MSIP grid &  Detectable g+r \\
SN 2018fdz & 18:15:18.54 & +29:54:38.69 & 0.065  & In MSIP grid &  Detectable g+r \\
SN 2018fdy & 00:53:13.01 & +20:42:52.95 & 0.082  & In MSIP grid &  Detectable g or r \\
SN 2018fdv & 00:14:44.78 & +17:52:05.69 & 0.094  & In MSIP grid &  Detectable g or r \\
SN 2018fcw & 00:50:55.08 & -07:46:00.98 & 0.050  & In MSIP grid &  No reference \\
SN 2018fcu & 04:17:54.69 & -49:54:28.48 & 0.050  & In MSIP grid &  No reference \\
SN 2018fcd & 15:40:11.09 & +61:31:48.05 & 0.085  & In MSIP grid &  Detectable g+r \\
SN 2018fbj & 22:19:58.95 & +40:04:22.60 & 0.099  & In MSIP grid &  No reference \\
SN 2018fbh & 15:36:31.98 & +41:47:59.63 & 0.041  & In MSIP grid &  Detectable g+r \\
SN 2018fae & 17:30:42.75 & +62:49:52.07 & 0.082  & In MSIP grid &  Detectable g+r \\
SN 2018ezz & 18:37:51.87 & +51:50:16.80 & 0.053  & In MSIP grid &  Detectable g+r \\
SN 2018ezy & 18:58:52.52 & +69:00:48.58 & 0.079  & In MSIP grid &  Detectable g+r \\
SN 2018ezx & 04:08:08.09 & -08:49:59.45 & 0.033  & In MSIP grid &  No reference \\
SN 2018eyh & 14:54:37.03 & +69:38:53.83 & 0.060  & In MSIP grid &  Detectable g+r \\
SN 2018exv & 23:33:57.44 & -26:26:49.28 & 0.070  & In MSIP grid &  No reference \\
SN 2018exh & 15:07:39.13 & +38:12:48.95 & 0.100  & In MSIP grid &  Detectable g+r \\
SN 2018exc & 21:00:08.02 & -40:21:30.94 & 0.057  & In MSIP grid &  No reference \\
SN 2018exb & 21:13:08.54 & -20:42:38.90 & 0.047  & In MSIP grid &  No reference \\
SN 2018evw & 21:15:14.40 & +02:11:34.44 & 0.050  & In MSIP grid &  No reference \\
SN 2018evt & 13:46:39.32 & -09:38:36.56 & 0.029  & In MSIP grid &  No reference \\
SN 2018evo & 21:48:38.42 & -43:22:48.07 & 0.077  & In MSIP grid &  No reference \\
SN 2018evf & 23:09:35.81 & +05:35:12.16 & 0.060  & In MSIP grid &  No reference \\
SN 2018evd & 23:16:51.85 & +41:03:33.72 & 0.060  & In MSIP grid &  No reference \\
SN 2018evc & 22:54:11.70 & +05:03:51.07 & 0.049  & In MSIP grid &  No reference \\
SN 2018evb & 22:51:02.72 & -03:38:58.15 & 0.090  & In MSIP grid &  No reference \\
SN 2018euz & 16:14:33.77 & +36:56:36.24 & 0.038  & In MSIP grid &  Detectable g+r \\
SN 2018euj & 19:52:29.60 & -60:45:51.40 & 0.034  & In MSIP grid &  No reference \\
SN 2018eui & 03:13:33.24 & -15:15:21.64 & 0.031  & In MSIP grid &  No reference \\
SN 2018etm & 16:14:20.68 & +03:13:55.08 & 0.020  & In MSIP grid &  Detectable g+r \\
SN 2018etj & 14:04:23.17 & +60:01:27.66 & 0.042  & In MSIP grid &  Detectable g+r \\
SN 2018esx & 01:12:12.47 & -30:29:40.07 & 0.090  & In MSIP grid &  No reference \\
SN 2018esq & 23:29:38.76 & -25:10:25.38 & 0.130  & In MSIP grid &  No reference \\
SN 2018esa & 22:41:02.82 & +20:43:32.02 & 0.070  & In MSIP grid &  Detectable g or r \\
SN 2018ery & 15:51:06.33 & +08:58:31.74 & 0.070  & In MSIP grid &  Detectable g or r \\
SN 2018ert & 16:35:22.54 & +22:28:06.23 & 0.094  & In MSIP grid &  Detectable g+r \\
SN 2018ers & 16:09:52.84 & +65:56:18.48 & 0.088  & In MSIP grid &  Detectable g+r \\
SN 2018err & 15:36:53.10 & +66:03:29.11 & 0.106  & In MSIP grid &  Detectable g+r \\
SN 2018ero & 20:08:41.22 & +03:16:17.60 & 0.060  & Outside MSIP grid & Not observable \\
SN 2018erl & 21:12:09.69 & -08:27:42.34 & 0.029  & In MSIP grid &  No reference \\
SN 2018erc & 02:07:28.51 & +11:09:13.64 & 0.041  & In MSIP grid &  No reference \\
SN 2018erb & 02:38:11.44 & +26:01:09.12 & 0.041  & In MSIP grid &  No reference \\
SN 2018eqq & 03:06:55.16 & +41:30:32.90 & 0.016  & In MSIP grid &  No reference \\
SN 2018eqg & 15:08:14.57 & +34:52:05.20 & 0.092  & In MSIP grid &  Detectable g+r \\
SN 2018epx & 23:29:53.22 & +27:22:40.90 & 0.024  & In MSIP grid &  No reference \\
SN 2018epw & 23:06:30.46 & +10:18:18.96 & 0.063  & In MSIP grid &  No reference \\
SN 2018eps & 16:20:43.18 & +65:38:20.96 & 0.070  & In MSIP grid &  Detectable g+r \\
SN 2018epj & 15:04:23.61 & +61:37:17.21 & 0.074  & In MSIP grid &  Detectable g+r \\
SN 2018epa & 21:51:16.88 & -14:27:36.22 & 0.040  & In MSIP grid &  No reference \\
SN 2018eoy & 20:39:28.80 & -13:07:58.31 & 0.070  & In MSIP grid &  No reference \\
SN 2018eod & 20:53:19.67 & -33:25:24.86 & 0.054  & Outside MSIP grid & Not observable \\
SN 2018enk & 00:25:51.83 & -20:40:50.34 & 0.050  & In MSIP grid &  No reference \\
SN 2018enj & 20:05:41.10 & -47:58:41.30 & 0.030  & In MSIP grid &  No reference \\
SN 2018eni & 18:19:11.78 & +54:05:19.29 & 0.085  & In MSIP grid &  Detectable g+r \\
SN 2018enc & 15:19:28.63 & -09:52:50.03 & 0.017  & Outside MSIP grid & Not observable \\
SN 2018enb & 21:39:34.57 & +08:52:47.30 & 0.023  & Chip / RC border & Not clear \\
SN 2018emy & 14:45:30.39 & +17:15:29.60 & 0.040  & In MSIP grid &  Detectable g+r \\
SN 2018emx & 15:26:58.79 & +08:15:41.37 & 0.088  & In MSIP grid &  Detectable g+r \\
SN 2018emv & 15:05:30.44 & +30:54:36.92 & 0.057  & In MSIP grid &  Detectable g+r \\
SN 2018emo & 01:05:38.12 & +29:29:50.82 & 0.067  & In MSIP grid &  No reference \\
SN 2018eml & 15:46:07.11 & +29:44:00.49 & 0.032  & In MSIP grid &  Detectable g+r \\
SN 2018emj & 16:00:03.58 & +23:22:50.62 & 0.100  & In MSIP grid &  Detectable g or r \\
SN 2018emi & 17:06:36.33 & +24:32:43.15 & 0.038  & In MSIP grid &  Detectable g+r \\
SN 2018emf & 00:31:38.41 & +28:58:10.50 & 0.072  & In MSIP grid &  No reference \\
SN 2018elm & 22:37:21.10 & +22:22:46.45 & 0.043  & In MSIP grid &  Detectable g+r \\
SN 2018ell & 16:49:57.03 & +27:38:26.94 & 0.064  & In MSIP grid &  Detectable g+r \\
SN 2018elj & 22:08:23.37 & +11:22:45.43 & 0.040  & In MSIP grid &  Detectable g+r \\
SN 2018elh & 21:06:09.64 & +70:21:08.71 & 0.050  & In MSIP grid &  No reference \\
SN 2018ekt & 16:11:28.40 & +45:27:11.70 & 0.015  & In MSIP grid &  Detectable g+r \\
SN 2018efm & 13:32:50.51 & +07:18:39.27 & 0.030  & In MSIP grid &  No reference \\
SN 2018efk & 12:22:36.82 & +47:58:09.35 & 0.050  & In MSIP grid &  Detectable g+r \\
SN 2018efe & 20:39:39.94 & +13:08:08.74 & 0.050  & In MSIP grid &  Detectable g+r \\
SN 2018efb & 17:44:44.31 & +38:09:50.54 & 0.100  & In MSIP grid &  Detectable g+r \\
SN 2018eew & 01:52:03.79 & -07:38:33.14 & 0.060  & In MSIP grid &  No reference \\
SN 2018eec & 22:39:33.10 & -02:45:41.23 & 0.090  & In MSIP grid &  No reference \\
SN 2018edz & 16:01:05.44 & +11:56:39.17 & 0.045  & In MSIP grid &  Detectable g or r \\
SN 2018edw & 19:08:40.56 & +78:28:39.78 & 0.080  & In MSIP grid &  Detectable g or r \\
SN 2018edd & 14:53:32.29 & +03:04:13.94 & 0.030  & In MSIP grid &  Detectable g or r \\
SN 2018ect & 22:35:08.40 & +17:23:20.66 & 0.070  & In MSIP grid &  Detectable g or r \\
SN 2018ecr & 18:23:04.99 & +27:28:16.03 & 0.070  & In MSIP grid &  Detectable g+r \\
SN 2018ecq & 16:45:42.76 & +37:53:25.54 & 0.100  & In MSIP grid &  Detectable g+r \\
SN 2018ecp & 15:30:21.23 & +27:37:46.26 & 0.070  & In MSIP grid &  Detectable g+r \\
SN 2018ecd & 13:19:36.73 & -29:07:16.61 & 0.050  & In MSIP grid &  No reference \\
SN 2018ebx & 00:55:26.42 & -74:18:42.19 & 0.034  & In MSIP grid &  No reference \\
SN 2018ebw & 02:41:16.40 & +20:30:16.88 & 0.030  & In MSIP grid &  No reference \\
SN 2018ebs & 21:39:37.62 & +08:56:39.78 & 0.030  & Chip / RC border & Not clear \\
SN 2018ebo & 14:16:48.50 & +58:29:07.45 & 0.080  & In MSIP grid &  Detectable g+r \\
SN 2018ebj & 03:53:06.79 & -45:10:51.28 & 0.051  & In MSIP grid &  No reference \\
SN 2018eaz & 18:42:55.21 & +50:39:33.60 & 0.050  & In MSIP grid &  Detectable g+r \\
SN 2018eak & 16:15:48.49 & +19:39:25.85 & 0.030  & In MSIP grid &  Detectable g or r \\
SN 2018eag & 18:57:16.32 & +46:29:56.75 & 0.072  & In MSIP grid &  Detectable g+r \\
SN 2018ead & 18:17:21.46 & +54:32:14.89 & 0.070  & In MSIP grid &  Detectable g+r \\
SN 2018dzy & 22:00:41.71 & +19:39:58.34 & 0.020  & In MSIP grid &  Detectable g or r \\
SN 2018dzr & 17:02:22.57 & +57:45:05.15 & 0.070  & In MSIP grid &  Detectable g+r \\
SN 2018dzh & 14:40:33.36 & +13:07:23.14 & 0.050  & Outside MSIP grid & Not observable \\
SN 2018dyz & 15:10:40.48 & +08:34:27.76 & 0.045  & In MSIP grid &  Detectable g or r \\
SN 2018dyq & 16:31:11.00 & +60:35:51.84 & 0.080  & In MSIP grid &  Detectable g+r \\
SN 2018dyp & 16:45:10.11 & +42:43:04.37 & 0.060  & In MSIP grid &  Detectable g+r \\
SN 2018dym & 16:05:04.16 & +36:05:38.71 & 0.090  & In MSIP grid &  Detectable g+r \\
SN 2018dyg & 14:27:12.79 & +16:51:45.61 & 0.050  & In MSIP grid &  Detectable g or r \\
SN 2018dye & 14:29:14.17 & +46:03:03.09 & 0.077  & In MSIP grid &  Detectable g+r \\
SN 2018dxu & 14:30:09.92 & +55:53:11.63 & 0.108  & In MSIP grid &  Detectable g+r \\
SN 2018dvf & 23:14:05.43 & +29:38:00.03 & 0.046  & In MSIP grid &  Detectable g+r \\
SN 2018dvd & 13:54:34.55 & +42:46:28.73 & 0.070  & In MSIP grid &  Detectable g+r \\
SN 2018dvb & 15:33:56.92 & +31:10:11.42 & 0.065  & In MSIP grid &  Detectable g+r \\
SN 2018dsw & 17:35:14.17 & +54:14:48.55 & 0.090  & In MSIP grid &  Detectable g+r \\
SN 2018dsv & 17:11:13.49 & +38:35:25.95 & 0.040  & Outside MSIP grid & Not observable \\
SN 2018dje & 17:52:50.55 & +21:22:57.60 & 0.040  & In MSIP grid &  Detectable g+r \\
SN 2018djd & 02:14:33.83 & -00:45:56.77 & 0.026  & Outside MSIP grid & Not observable \\
SN 2018dhw & 18:59:57.52 & +72:16:02.15 & 0.029  & In MSIP grid &  Detectable g+r \\
SN 2018dgz & 15:45:41.62 & +11:57:09.73 & 0.070  & In MSIP grid &  Detectable g or r \\
SN 2018dgk & 13:52:57.96 & +05:17:23.18 & 0.080  & In MSIP grid &  Detectable g or r \\
SN 2018des & 17:20:08.95 & +09:29:32.25 & 0.080  & In MSIP grid &  Detectable g or r \\
SN 2018der & 14:48:05.62 & +63:13:00.03 & 0.052  & In MSIP grid &  Detectable g+r \\
SN 2018deq & 18:11:57.33 & +30:02:43.50 & 0.067  & In MSIP grid &  Detectable g+r \\
SN 2018dem & 21:11:58.53 & -00:13:05.16 & nan  & In MSIP grid &  No reference \\
SN 2018dej & 21:09:46.14 & -50:14:12.41 & 0.058  & In MSIP grid &  No reference \\
SN 2018dei & 13:23:43.12 & -25:24:32.33 & 0.041  & In MSIP grid &  No reference \\
SN 2018ddy & 22:59:57.34 & -45:25:53.33 & 0.051  & In MSIP grid &  No reference \\
SN 2018dds & 17:44:18.15 & +68:01:44.66 & 0.075  & In MSIP grid &  Detectable g+r \\
SN 2018ddi & 03:39:35.12 & -06:19:30.29 & 0.021  & In MSIP grid &  No reference \\
SN 2018ddh & 12:18:44.04 & +44:46:55.11 & 0.038  & In MSIP grid &  Detectable g+r \\
SN 2018ddg & 14:12:18.25 & +62:38:34.72 & 0.073  & In MSIP grid &  Detectable g+r \\
SN 2018ddb & 00:02:49.71 & -66:11:06.14 & 0.029  & In MSIP grid &  No reference \\
SN 2018dda & 22:08:14.15 & -25:03:41.58 & 0.018  & In MSIP grid &  No reference \\
SN 2018dcn & 17:06:44.02 & +18:21:25.41 & 0.050  & In MSIP grid &  Detectable g or r \\
SN 2018dcm & 17:25:41.47 & +23:52:08.96 & 0.064  & In MSIP grid &  Detectable g+r \\
SN 2018dbe & 14:27:55.24 & +40:58:27.64 & 0.087  & In MSIP grid &  Detectable g+r \\
SN 2018dbd & 16:52:47.35 & +51:33:48.34 & 0.075  & In MSIP grid &  Detectable g+r \\
SN 2018dbc & 15:21:24.31 & +69:36:22.63 & 0.080  & In MSIP grid &  Detectable g+r \\
SN 2018cxm & 14:53:58.33 & +26:00:07.09 & 0.048  & In MSIP grid &  Detectable g+r \\
SN 2018cxl & 17:44:24.63 & +55:10:31.56 & 0.060  & In MSIP grid &  Detectable g+r \\
SN 2018cxj & 12:46:30.22 & +77:16:51.51 & 0.050  & In MSIP grid &  Detectable g or r \\
SN 2018cxe & 17:21:01.63 & +26:09:07.31 & 0.044  & Outside MSIP grid & Not observable \\
SN 2018cvx & 12:51:27.49 & +20:36:23.29 & 0.062  & In MSIP grid &  Detectable g+r \\
SN 2018cvw & 15:57:02.31 & +37:25:01.84 & 0.031  & In MSIP grid &  Detectable g+r \\
SN 2018cvv & 12:04:35.87 & +12:33:24.45 & 0.065  & In MSIP grid &  Detectable g or r \\
SN 2018cvu & 13:15:34.32 & +28:20:34.48 & 0.102  & In MSIP grid &  Detectable g+r \\
SN 2018cvt & 14:51:53.63 & +46:38:44.70 & 0.074  & In MSIP grid &  Detectable g+r \\
SN 2018cvs & 13:41:07.97 & +28:53:11.47 & 0.105  & In MSIP grid &  Detectable g+r \\
SN 2018cvr & 14:45:24.55 & +78:17:05.00 & 0.100  & Chip / RC border & Not clear \\
SN 2018cvq & 15:52:54.43 & +50:37:10.10 & 0.067  & In MSIP grid &  Detectable g+r \\
SN 2018cvh & 15:07:58.60 & +01:13:56.63 & 0.035  & In MSIP grid &  No reference \\
SN 2018cvf & 12:06:40.62 & +59:30:48.11 & 0.064  & In MSIP grid &  Detectable g+r \\
SN 2018cvd & 13:14:40.26 & +24:00:20.68 & 0.067  & In MSIP grid &  Detectable g+r \\
SN 2018cuw & 18:46:14.38 & +35:58:07.27 & 0.024  & In MSIP grid &  Detectable g+r \\
SN 2018cua & 15:32:03.56 & +23:31:01.33 & 0.091  & In MSIP grid &  Detectable g+r \\
SN 2018cty & 13:01:45.70 & +61:27:55.73 & 0.056  & In MSIP grid &  Detectable g+r \\
SN 2018cts & 13:59:14.00 & +28:32:26.76 & 0.064  & In MSIP grid &  Detectable g or r \\
SN 2018ctq & 12:13:47.91 & +40:42:56.35 & 0.100  & In MSIP grid &  Detectable g+r \\
SN 2018cto & 17:24:32.04 & +70:22:20.81 & 0.051  & In MSIP grid &  Detectable g+r \\
SN 2018ctm & 18:26:14.18 & +45:00:36.69 & 0.065  & In MSIP grid &  Detectable g+r \\
SN 2018cti & 22:17:30.73 & +11:43:17.58 & 0.035  & Outside MSIP grid & Not observable \\
SN 2018ctc & 14:18:18.02 & +28:58:23.88 & 0.042  & In MSIP grid &  Detectable g or r \\
SN 2018ctb & 22:01:08.36 & +20:03:05.29 & 0.029  & In MSIP grid &  No reference \\
SN 2018crs & 17:20:24.40 & +55:12:52.72 & 0.072  & In MSIP grid &  Detectable g+r \\
SN 2018crr & 16:59:37.23 & +59:04:23.73 & 0.073  & In MSIP grid &  Detectable g+r \\
SN 2018cro & 21:00:49.26 & +09:31:28.18 & 0.076  & In MSIP grid &  No reference \\
SN 2018crn & 18:55:21.84 & +56:35:18.14 & 0.050  & In MSIP grid &  Detectable g+r \\
SN 2018crj & 17:58:02.75 & +69:04:22.13 & 0.088  & In MSIP grid &  Detectable g+r \\
SN 2018cri & 16:11:21.47 & +36:59:39.41 & 0.064  & In MSIP grid &  Detectable g+r \\
SN 2018cqw & 18:17:32.21 & +19:26:40.49 & nan  & Outside MSIP grid & Not observable \\
SN 2018cqv & 23:18:28.71 & -25:59:25.38 & 0.103  & In MSIP grid &  No reference \\
SN 2018cqj & 09:40:21.46 & -06:59:19.76 & 0.021  & In MSIP grid &  No reference \\
SN 2018cqa & 14:47:45.78 & +32:45:06.80 & 0.060  & Outside MSIP grid & Not observable \\
SN 2018coy & 20:20:28.88 & -33:23:37.00 & 0.039  & Outside MSIP grid & Not observable \\
SN 2018cov & 23:49:48.24 & -14:07:09.95 & 0.051  & In MSIP grid &  No reference \\
SN 2018coj & 12:01:54.86 & +36:53:42.71 & 0.079  & In MSIP grid &  Detectable g or r \\
SN 2018coi & 16:18:57.98 & +56:43:00.77 & 0.059  & In MSIP grid &  Detectable g+r \\
SN 2018coh & 17:20:38.66 & +52:02:30.97 & 0.085  & In MSIP grid &  Detectable g+r \\
SN 2018cof & 14:59:54.39 & +39:04:43.78 & 0.092  & In MSIP grid &  Detectable g+r \\
SN 2018coe & 16:51:37.37 & +61:32:43.34 & 0.080  & In MSIP grid &  Detectable g+r \\
SN 2018cod & 13:20:27.96 & +62:18:03.02 & 0.030  & In MSIP grid &  Detectable g+r \\
SN 2018coc & 16:38:32.00 & +05:07:35.30 & 0.090  & In MSIP grid &  Detectable g or r \\
SN 2018cny & 15:04:16.62 & +35:48:54.11 & 0.047  & In MSIP grid &  Detectable g+r \\
SN 2018cnu & 02:10:02.21 & +37:02:18.31 & 0.025  & In MSIP grid &  No reference \\
SN 2018cnp & 13:49:39.47 & +47:49:09.37 & 0.028  & Outside MSIP grid & Not observable \\
SN 2018cno & 02:10:32.76 & -06:52:27.37 & 0.043  & In MSIP grid &  No reference \\
SN 2018cng & 15:45:29.05 & +35:51:19.09 & 0.066  & In MSIP grid &  Detectable g+r \\
SN 2018cne & 16:18:42.70 & +40:04:20.83 & 0.080  & In MSIP grid &  Detectable g+r \\
SN 2018cmu & 09:17:59.90 & +50:00:07.81 & 0.034  & In MSIP grid &  Detectable g or r \\
SN 2018cmt & 23:56:55.72 & -24:59:47.58 & 0.074  & In MSIP grid &  No reference \\
SN 2018cmo & 23:14:22.23 & -02:02:01.86 & 0.026  & In MSIP grid &  No reference \\
SN 2018cmk & 11:38:29.86 & +20:31:44.26 & nan  & In MSIP grid &  Detectable g or r \\
SN 2018cjy & 12:59:45.30 & -25:36:06.13 & 0.064  & In MSIP grid &  No reference \\
SN 2018cjw & 17:46:41.41 & +27:35:22.93 & 0.094  & In MSIP grid &  Detectable g+r \\
SN 2018cjn & 18:46:20.42 & +70:44:13.46 & 0.100  & In MSIP grid &  Detectable g or r \\
SN 2018cjg & 13:31:55.35 & +23:16:59.85 & 0.045  & In MSIP grid &  Detectable g or r \\
SN 2018cjd & 14:58:20.02 & -37:33:25.16 & 0.026  & In MSIP grid &  No reference \\
SN 2018cif & 22:03:00.93 & +02:35:51.31 & 0.029  & Outside MSIP grid & Not observable \\
SN 2018cfa & 16:49:39.12 & +45:29:32.64 & 0.030  & In MSIP grid &  Detectable g+r \\